\documentclass[preprint,preprintnumbers,amsmath,amssymb]{revtex4}

\usepackage{etex}
\usepackage{amssymb,amsthm,amscd,amsbsy,array}
\usepackage{bm}
\usepackage{soul} 
\usepackage{graphics,graphicx,xcolor}
\usepackage[colorlinks=true, pdfstartview=FitV, linkcolor=blue, citecolor=blue, urlcolor=blue]{hyperref}
\numberwithin{equation}{section}
\renewcommand{\theequation}{\thesection.\arabic{equation}}
\let\ssection=\section
\renewcommand{\section}{\setcounter{equation}{0}\ssection}
\newcommand{\bA}{{\bf A}}

\newcommand{\bbgamma}{\boldsymbol{\gamma}}

\newcommand{\Ad}{\mathrm{Ad}}

\newcommand{\bb}{{\bf b}}
\newcommand{\vB}{{\mathbf{B}}}
\newcommand{\vE}{{\mathbf{E}}}
\newcommand{\vD}{{\mathbf{D}}}
\newcommand{\vH}{{\mathbf{H}}}
\newcommand{\vj}{{\mathbf{j}}}
\newcommand{\bbeta}{\boldsymbol{\beta}}

\newcommand{\tbB}{\widetilde{\vB}}

\newcommand{\bc}{{\mathbf{c}}}

\newcommand{\Coad}{\mathrm{Coad}}

\newcommand{\Carr}{{\mathrm{Carr}}}
\newcommand{\carr}{{\mathfrak{carr}}}

\def\Barg{{{\rm Barg}}}
\newcommand{\barg}{\mathfrak{barg}}
\newcommand{\Gal}{\mathrm{Gal}}
\newcommand{\gal}{\mathfrak{gal}}

\newcommand{\Div}{\mathrm{Div}}

\newcommand{\rE}{{\mathrm{E}}}

\newcommand{\trE}{\widetilde{\rE}}
\newcommand{\tbE}{\widetilde{\vE}}

\newcommand{\cF}{{\mathcal{F}}}

\newcommand{\hcF}{\widehat{\mathcal{F}}}

\newcommand{\bg}{{\bf g}}

\newcommand{\rg}{\mathrm{g}}
\newcommand{\BG}{{G}}

\newcommand{\fg}{\mathfrak{g}}

\newcommand{\bgamma}{\boldsymbol{\gamma}}

\newcommand{\rO}{{\mathrm{O}}}

\newcommand{\cO}{{\mathcal{O}}}

\newcommand{\bomega}{{\boldsymbol{\omega}}}

\newcommand{\bsell}{{\boldsymbol{\ell}}}

\newcommand{\bp}{{\bf p}}

\newcommand{\bq}{{\bf q}}

\newcommand{\bx}{{\bm{x}}}

\newcommand{\bbR}{\mathbb{R}}

\newcommand{\grad}{{\vnabla}}
\newcommand{\rot}{\vnabla\times}

\newcommand{\dive}{{\vnabla\cdot}}

\newcommand{\bs}{{\bf s}}

\newcommand{\soo}{\mathfrak{so}}

\newcommand{\Tr}{\mathrm{Tr}}

\newcommand{\bv}{{\bf v}}

\def\vnabla{{\bm{\nabla}}}
\def\bnabla{{\bm{\nabla}}}

\def\Ort{{\rm O}}

\def\vbeta{{\bm{\beta}}}
\def\vb{{\bm{b}}}

\def\bv{{\bm{v}}}
\def\bp{{\bm{p}}}

\def\bq{{\bm{q}}}

\def\beq{\begin{equation}}
\def\eeq{\end{equation}}
\def\beqa{\begin{eqnarray}}
\def\eeqa{\end{eqnarray}}
\def\nn{\nonumber}
\def\barray{\left(\begin{array}}
\def\earray{\end{array}\right)}
\def\barraynb{\begin{array}}
\def\earraynb{\end{array}}

\def\Ort{{\rm O}}
\def\IR{{\mathbb{R}}} 

\def\IC{{{C}}} 
\def\IN{{{N}}} 
\def\IB{{{B}}} 
\def\IE{{{E}}}

\def\?{\quad{\gb{\fbox{\texttt{?}}\;}}\quad}
\def\p{{\partial}}

\def\vv{\mathbf{v}}
\def\vx{\mathbf{x}}
\def\v0{\mathbf{0}}

\usepackage{color}

\newcommand{\gb}{\colorbox{green}}

\def\beq{\begin{equation}}
\def\eeq{\end{equation}}
\def\bea{\begin{eqnarray}}
\def\eea{\end{eqnarray}}

\def\p{\partial}

\def \p{{\partial}}

\newcommand{\s}{\sigma}

\def\6{\partial}
\def\7{\tilde}\def\t{\widetilde}
\def\8{\widehat}
 \def\bx{{\bf x}}

\def\G11{\Gamma_{11} }

\newcommand{\vM}{{\mathbf{M}}}
\newcommand{\vN}{{\mathbf{N}}}

\newcommand{\const}{\mathop{\rm const.}\nolimits}
\newcommand{\half }{\frac{1}{2}}

\renewcommand{\theequation}{\thesection.\arabic{equation}}
\let\ssection=\section
\renewcommand{\section}{\setcounter{equation}{0}\ssection}

\begin{document}

\preprint{arXiv:1402.0657v5 [gr-qc]
}

\title{Carroll versus Newton
and Galilei:\\
two dual non-Einsteinian concepts of time
\\[6pt]
}

\author{
C. Duval$^{1}$\footnote{Aix-Marseille Universit\'e, CNRS, CPT, UMR 7332, 13288 Marseille, France.
Universit\'e de Toulon, CNRS, CPT, UMR 7332, 83957 La Garde, France.
mailto:duval@cpt.univ-mrs.fr},
G. W. Gibbons$^{2}$\footnote{
mailto:G.W.Gibbons@damtp.cam.ac.uk},
P. A. Horvathy$^{3,4}$\footnote{mailto:horvathy@lmpt.univ-tours.fr},
P. M. Zhang$^{3}$\footnote{mailto:zhpm@impcas.ac.cn}
}

\affiliation{
$^1$Centre de Physique Th\'eorique,
Marseille, France
\\
$^2$D.A.M.T.P., Cambridge University, U.K.
\\
$^3$Institute of Modern Physics, Chinese Academy of Sciences,
Lanzhou, China
\\
$^4$Laboratoire de Math\'ematiques et de Physique
Th\'eorique,
Universit\'e de Tours,
France
}

\date{\today}

\begin{abstract}
The Carroll group was  originally
introduced by L\'evy-Leblond [1] by considering the contraction of the Poincar\'e
group as  $c\to0$.
In this paper an alternative  definition, based on
the geometric properties of a non-Minkowskian, non-Galilean
but nevertheless boost-invariant,  space-time structure is proposed.
A ``duality'' with the Galilean limit $c\to\infty$ is established.
Our theory is illustrated by Carrollian electromagnetism.
\end{abstract}

\maketitle

\baselineskip=16pt

\tableofcontents

\newpage

\section{Introduction}\label{Intro}

The last decade or so has  witnessed  an increased  interest in
non-Minkowskian spacetimes whose structures are   nevertheless
 \emph{invariant under boosts}.

These   may  be obtained by group contraction.
The standard  contraction
of the Poincar\'e group yields  the Galilei group  \cite{Inonu},
for  which pure Lorentz transformations become Galilei boosts.
However what is less well-known is  another rather unfamiliar
limit which yields instead a different
but still boost-invariant theory \cite{Leblond,SenGupta}.
L\'evy-Leblond \cite{Leblond}, who
introduced  this ``degenerate cousin of
the Poincar\'e group'',
named  it, with  tongue-in-cheek,
the \emph{Carroll group}, referring to the pseudonym of the author of \textit{Through the Looking-Glass
 } \cite{Alice}.

Now, as we shall recall in detail later,
the Galilei  and Carroll groups are both
related by a fascinating type of \textit{``duality''},
namely one between  two different sorts of ``times'' we denote by $t$ and $s$, respectively.

The quantity $t$ is the absolute time of Newton \cite{Newton}. In his own words :
\begin{quote}\textit{\narrower Absolute, true, and mathematical time,
in and of itself and of its own nature, without reference to anything external, flows uniformly and by another name is called duration}~\dots
\end{quote}
while $s$  is Carrollian  \cite{Alice} ``time''. In the words of the Red Queen:
\begin{quote}
\textit{\narrower
    ``Well, in our country,'' said Alice, still panting a little,
``you'd generally get to somewhere else if
you run very fast for a long time, as we've been doing.''}

 \textit{\narrower  ``A slow sort of country!" said the Queen. ``Now, here, you see, it takes all the running you can do, to keep in the same place. If you want to get somewhere else, you must run at least twice as fast as that!''
}
\end{quote}

Setting aside philosophical reflections,
the Galilei and the Carroll groups will turn out  to be the space-time symmetries of two \textit{different} types
of   $(d+1)$-dimensional  non-Minkowskian spacetimes we call $\IN$ and $\IC$, respectively,
upon which ``boosts'' act --- but they act differently.

We use here the adjective \emph{non-Minkowskian} deliberately.
Galilean, or  Newton-Cartan spacetime is
often referred to as a ``non-relativistic''
spacetime in contrast to   Minkowski spacetime which is referred to
as ``relativistic''. However both usages ignore the fact that
both Galilean physics and Einsteinian  physics
admit a \emph{relativity principle}, and have ``boosts'' as part of their underlying
symmetry. It is just that those ``boosts'' do not act in the same way. Einstein's
great achievement was to see that  both principles could not be
simultaneously  true in any consistent theory and to decide which one should be abandoned.

 In what follows we shall be defining
curved Newton-Cartan and curved  Carroll spacetimes modelled on
their flat versions. By analogy with the customary terminology
in General Relativity where one introduces curved Lorentzian spacetimes
modelled on flat Minkowski spacetime we shall refer to them as non-Lorentzian
spacetimes.

Although neither of our two  spacetimes is Minkowskian,
both $\IN$ and $\IC$ may be obtained in a unified fashion, namely from a
$(d+2)$-dimensional {\sl Minkowski}   space-time, $\IE^{d+1,1}$  \cite{DBKP,DGH}. Moreover, the
duality we are interested in is
 best seen, in our view, from this higher dimensional perspective.

From a geometrical point of view,
the duality  is between \emph{pushforward} and \emph{pullback}.
Thus  $\IN$ is   a  Kaluza-Klein-type
``lightlike shadow'', i.e., the quotient space of $\rE^{d+1,1}$ by a group of null translation,
 while $\IC$  may be obtained as an embedded
 lightlike $(d+1)$-brane, i.e., a null hyperplane of Minkowski space $\rE^{d+1,1}$.

Having at our disposal two null coordinates
$t$ and $s$, in $\rE^{d+1,1}$, our hyperplane $C$ will be given by a slice $t=\const$, while
$s$ will parameterize null translations.

Then the  duality we are alluding to
consists of the \emph{reflection swapping the light-cone coordinates $s$ and $t$}
\cite{HHAP,DLazzari}.
Since $t$ and $s$ play the roles of  time coordinates in $\IN$ and $\IC$
respectively, the duality is associated with
\emph{two  different non-Minkowskian notions of ``time''}.

\goodbreak

 The potential applications of our results include the possibility
of developing a notion of  holography for asymptotically
flat space-times, a primary motivation for much  recent
\cite{Bag2,Barnich1,Arcioni,Schroer}
and indeed older  \cite{Henneaux1} work in quantum gravity
(see \cite{Melas}) for a review).

 The  Carroll group also emerges naturally
in brane-dynamics in  the limit that the brane world volume becomes
lightlike \cite{GiHaYi,thoughts}.
The possible role of Carrollian space-times
near space-time singularities and in the so called strong coupling
(i.e. small  gravitational tension   $c^4/G$ \cite{Max}) limit of
General Relativity has been discussed by Henneaux \cite{Henneaux1}
and by Dautcourt \cite{Dautcourt}
and this is closely related to   work on Kac-Moody symmetries
in M-theory \cite{Nicolai}.

Another
potential application of our results is to the quantization of
quantum fields on null hypersurfaces, such as causal  horizons
located in the interior of a dynamical space-time, cf. \cite{Wall1,Wall2}.
It is also worth mentioning the recent study of the relation between electric-magnetic duality with Poincar\'e invariance \cite{BunsterHe}.

The organization of the paper is as follows.
After recalling the construction of the Carroll group~\cite{Leblond,SenGupta} by \emph{contraction
of  the Poincar\'e group}, we propose two
further, alternative definitions.

First, we define the Carroll group directly, as associated with the proper structure of non-relativistic space-time, with no reference to
relativistic ancestry.
Intuitively, the Carroll group is for
a Carroll structure as Galilean isometries
are for a Newton-Cartan structure
\cite{GalIso,DHGalConf}.

Then we show that the Carroll group can, in fact,
be viewed as a \emph{subgroup of $E(d+1,1)$, the
Poincar\'e group in $(d+1,1)$ dimension}. Our construction here is then analogous to the geometric definition of the ``Bargmann''
[i.e., the mass-centrally extended Galilei] group, see \cite{DBKP,DGH}.

Then we illustrate our theory with some mathematical and physical examples.
We will study, in particular,  aspects of
non-Einsteinian electrodynamics, along the lines indicated by Le Bellac and L\'evy-Leblond \cite{LBLL}, and of others
\cite{SourElec,Kun2,GiHaYi,GoSh,Rousseaux}.

Another example is provided by the Chaplygin gas \cite{BaJa}, \cite{HHAP}.

Classical elementary models for the Carroll group can also be constructed \cite{Gomis,Ancille},
 but those turn out to be rather disappointing, since free Carroll particles \dots cannot  move.

\section{The Carroll group as a contraction}

The Carroll group was first constructed as a novel type contraction of the Poincar\'e group, $\rE(d,1)$, in \cite{Leblond,SenGupta}.

 Let us start with reminding the Reader of how the familiar Galilean limit is obtained.
Denoting by $x^0, x^A$ the affine Lorentzian coordinates on Minkowski spacetime $E^{d,1}$ \footnote{Roman indices $A,B,\ldots$  run from $1$ to $d$ and  Einstein summation  is understood.}, the {\emph{covariant}} metric reads
\beq
G = -dx^0\otimes dx^0 + \delta_{AB}\, dx^A \otimes dx^B.
\label{MinkG}
\eeq
Then the defining the time coordinate by
\beq
t=x^0/c
\label{Galtime}
\eeq
(where $c$ denotes the speed of light),
in term of which the \emph{contravariant} metric (or co-metric) is
\begin{equation}
G^{-1} =
- \frac{1}{c^2}
\frac{\partial}{\partial{}t} \otimes
\frac{\partial}{\partial{t}}
+
\delta^{AB}\frac{\partial}{\partial x^A} \otimes
\frac{\partial}{\partial x^B}.
\label{GalLim}
\end{equation}
Then the Galilean limit is obtained by
 letting $c\uparrow \infty$, when (\ref{GalLim})
becomes degenerate,
\beq
G^{-1}\rightarrow
\frac{\partial}{\partial x^A} \otimes
\frac{\partial}{\partial x^A},
\label{Galilei}
\eeq
with  kernel generated by the
co-normals $dt$ to the surfaces of constant time.

This limiting procedure should be contrasted with the one put forward by L\'evy-Leblond in Ref. \cite{Leblond}, who suggested to consider another ``time'' we denote here by $s$,
\begin{equation}
s=C x^0
\label{Carrtime}
\end{equation}
for some \emph{new constant} $C$ which has, once again, the dimension of velocity, and is designed so that the novel ``time'' coordinate, $s$, has actually the dimension of a (squared length)/time, $[s]=L^2T^{-1}$, i.e., an \textit{action/mass}.
 The Minkowski metric (\ref{MinkG}) is written,  in these new coordinates,
\beq
G=-\frac1{C^2}\,ds\otimes{}ds+\delta_{AB}\, dx^A \otimes dx^B
\eeq
 so that the \emph{Carrollian limit $C\uparrow\infty$} can now be considered, yielding another degenerate metric, namely
\beq
G\rightarrow\delta_{AB}\, dx^A \otimes dx^B.
\label{Carrollmet}
\eeq
whose  kernel is given by the direction of
$\partial/\partial s$.
A manifold with such a metric will be called a \emph{Carrollian space-time} and denoted by $\IC^{d+1}$.

The \emph{Carroll group}, denoted by $\Carr(d+1)$, is then obtained from the orthochronous Poincar\'e group, $E_+(d,1)$, by a contraction $C\uparrow{\infty}$.  In detail,
let us consider a Lorentz boost of Minkowski space-time defined by the dimensionless $3$-vector~$\vbeta$, namely
\beq
\left\{\barraynb{lll}
\vx'&=&\displaystyle\vx+(\gamma-1)\frac{(\vbeta\cdot\vx)\vbeta}{\vbeta^2}+
\vbeta\gamma\,x^0,
\\[8pt]
x'{}^0&=&\gamma(x^0+\vbeta\cdot\vx),
\earraynb\right.
\label{Lorenteztraf}
\eeq
where $\gamma=(1-\vbeta^2)^{-\half}$.
Putting now
\beq
\vb
=-C\vbeta,
\label{carlim}
\eeq
where the minus sign has been chosen for further convenience, we end up, in the limit $C\uparrow\infty$ (where $\vx,s,\vb$ are fixed), with the \textit{Carrollian boosts}
\beq
\left\{\barraynb{lll}
\vx'&=&\vx\\
s'&=&s-\vb\cdot\vx,
\earraynb\right.
\label{carrboost}
\eeq
with $\vb\in\bbR^3$.
The Carrollian limit $C\uparrow\infty$ of relativistic time-translations:
$\bx'=\bx$, and ${x^0}'= x^0+a^0$, yields in turn
\begin{equation}
\left\{
\barraynb{lll}
\vx'&=&\vx,\\
s'&=&s+f
\earraynb\right.
\label{carttr}
\end{equation}
with Carrollian \textit{``time''-translations}
$
f=Ca^0.
$

For the sake of comparison, we mention that introducing, as usual, the time coordinate $t=x^0/c$, and considering instead
$
\vb=c\,\vbeta,
$
 would provide us, in the limit $c\uparrow\infty$, with ordinary \textit{Galilei boosts},
\beq
\left\{\barraynb{lll}
\vx'&=&\vx+\vb\,t
\\
t'&=&t
\earraynb\right.
\label{Galboost}
\eeq
with $\bb\in\IR^3$.

Let us emphasise that $t$ and $s$
 in (\ref{Galtime}) and  in  (\ref{Carrtime}), respectively,
are different [non-Minkowskian] ``times'', in that they have \emph{different physical dimensions}.

The Carroll group $\Carr(d+1)$, i.e., the $C\uparrow\infty$-contraction of $\rE_+(d,1)$ is generated by boosts (\ref{carrboost}), orthogonal transformations: $\bx'=R\,\bx$, and $s'=s$ with $R\in\rO(d)$, space-translations which are clearly not affected by the contraction procedure, as well as ``time''-translations (\ref{carttr}).
As we shall see below (Eq. (\ref{CarrGlobal})), the Carroll group is the semi-direct product,
$\Carr(d+1) = \trE(d)\ltimes\bbR^{d+1}$,
of a group $\trE(d)$ --- isomorphic to the Euclidean group $\rE(d)$ --- and of the additive group $\bbR^{d+1}$ (described by the pairs $(\bx,s)$), and interpreted as Carroll spacetime.
See Eq. (\ref{Carr}), capturing the global structure of the Carroll group, $\Carr(d+1)$.
See also \cite{Huang}.

\section{Carroll structures: geometrical definition}
\label{CarrollSection}

Let us  now present a general geometric definition of \emph{Carroll manifolds and transformations} which makes no mention of the Poincar\'e,
group and its contractions but rather
is \textit{dual} to that of Newton-Cartan manifolds   \cite{GalIso,DHGalConf}.
In order to motivate our definitions in Sec. \ref{Carrollmanif}, we first remind the reader  of that case.

\subsection{Newton-Cartan manifolds}

A Newton-Cartan (NC) manifold is a quadruple
$
(\IN,\gamma,\theta,\nabla),
$
where $\IN$ (for Newton) is a smooth
$(d+1)$-dimensional manifold, $\gamma$ a twice-symmetric, contra\-variant, positive tensor field, whose kernel is generated by the nowhere vanishing  $1$-form $\theta$. Moreover, $\nabla$ is a symmetric affine con\-nection that parallel-transports both $\gamma$ and $\theta$ \footnote{Extra conditions on the symmetries of the curvature tensor of $\nabla$ are usually imposed to the definition (we will not really need them here); see \cite{DK}.
 Note that, even with these additional conditions, $\nabla$ is not uniquely defined by the ``Galilei structure'' $(N,\gamma,\theta)$.}.
See the comprehensive Reference \cite{Kunzle}.

The ``clock'' one-form $\theta$ is closed, $d\theta=0$, thus $\ker\theta$ is a Fr\"obenius-integrable dis\-tribution, whose leaves  are $d$-dimensional and are endowed with a Riemannian structure inherited from $\gamma$ \cite{Kunzle}. The quotient
 $K=\IN/\ker\theta$ (``Kronos'')
is $1$-dimensional: it is the absolute Newtonian time-axis (either compact or non-compact).

The standard flat NC structure is given, in an adapted coordinate system, by
\begin{equation}
\IN^{d+1}=\bbR\times\bbR^d,
\qquad
\gamma=\delta^{AB}\frac{\partial}{\partial{}x^A}\otimes\frac{\partial}{\partial{}x^B},
\qquad
\theta=dt,
\qquad
\Gamma^k_{ij}=0
\label{NC}
\end{equation}
for all $i,j,k={0,1,\ldots,d}$, and where $t=x^0$ is the Galilean time-coordinate.
See \cite{DHGalConf} for other non-trivial NC structures.
\goodbreak

The automorphisms [i.e., transformations which preserve \emph{all} geometrical in\-gredients $\gamma$, $\theta$, and~$\nabla$ of the theory] of the flat NC (Newton-Cartan) structure (\ref{NC}) constitute the \textit{Galilei group}, $\Gal(d+1)$, represented by the  matrices \cite{Sou,LLcoho}
\begin{equation}
g =
\left(
\begin{array}{ccc}
R&\bb&\bc\\
0&1&e\\
0&0&1
\end{array}
\right)\in\Gal(d+1),
\label{Gal}
\end{equation}
where $R\in\rO(d)$, $\bb,\bc\in\bbR^d$, and $e\in\bbR$ represent  orthogonal transformations, boosts, space and time translations, respectively. Cf. \cite{Trautman}.
 Then the Galilei Lie algebra $\gal(d+1)$ is isomorphic to the Lie algebra of vector fields on $N$,
\begin{equation}
X=(\omega^A_B\,x^B+\beta^A t +\gamma^A)\frac{\partial}{\partial{}x^A}+\varepsilon\frac{\partial}{\partial{}t}\in\gal(d+1),
\label{gal}
\end{equation}
where $\omega\in\soo(d)$, $\bbeta, \bbgamma\in\IR^d$, and $\varepsilon\in\IR$.
The action of $g\in\Gal(d+1)$ on Galilei space-``time'', $N$, thus reads
\begin{equation}
g_N:\left(\begin{matrix}
\bx\\t
\end{matrix}
\right)
\mapsto
\left(\begin{matrix}
R\,\bx+\bb t+\bc\\
t+e
\end{matrix}
\right)
\label{GalAction}
\end{equation}
for all $\bx\in\IR^d$, and $t\in\IR$.

Let us mention, \textit{en passant}, that the homogeneous Galilei group generated by orthogonal transformations and boosts is isomorphic to the Euclidean group in $d$ dimensions, spanned the matrices
\begin{equation}
h =
\left(
\begin{array}{cc}
R&\bb\\
0&1
\end{array}
\right)\in\rE(d).
\label{homGal}
\end{equation}
\goodbreak
\subsection{Carroll manifolds}\label{Carrollmanif}

 Now we propose an analogous definition of a
\emph{Carroll manifold} given by a quadruple $(\IC,\rg,\xi,\nabla)$,
where $C$ (for Carroll) is again a smooth $(d+1)$-dimensional manifold, endowed with a twice-symmetric covariant, positive, tensor field $\rg$, whose kernel is generated by the nowhere vanishing, complete vector field $\xi$, and $\nabla$ is a sym\-metric affine con\-nection that parallel-transports both $\rg$ and $\xi$. Note that, just as in the Galilei framework, the degeneracy of the ``metric'' $\rg$ implies that the con\-nection~$\nabla$ is not uniquely defined by the pair~$(\rg,\xi)$.

The group of automorphisms of a Carroll structure
 will be called the \textit{Carroll group} and denoted by $\Carr(C,\rg,\xi,\nabla)$. It consists of all diffeomorphisms of $C$ that preserve the ``metric'' $\rg$, the vector field $\xi$, as well as the connection $\nabla$.
The \textit{Carroll Lie algebra}, $\carr(C,\rg,\xi)$, is then identified with the Lie algebra of those vector fields~$X$ of $C$ such that
\begin{equation}
L_X\rg=0,
\qquad
L_X\xi=0,
\qquad
L_X\nabla=0.
\label{carrC}
\end{equation}
\goodbreak

The standard flat Carroll structure is given, in an adapted coordinate system, by
\begin{equation}
C^{d+1}=\bbR\times\bbR^d,
\qquad
\rg=\delta_{AB}\,dx^A\otimes{}dx^B,
\qquad
\xi=\frac{\partial}{\partial s},
\qquad
\Gamma^k_{ij}=0
\label{flatC}
\end{equation}
for all $i,j,k=0,1,\ldots,d$, where $s=x^0$ is now the ``Carrollian time'' coordinate. The coordinate~$s$ has the dimension of an \textit{action per mass}, in accordance with Eq. (\ref{Carrtime}); this will also be corroborated by the canonical construction of Carroll structures in terms of Bargmann ones as elaborated in Section \ref{CarrInBargSection}. See also the form of the flat Bargmann metric, $G$, in Eqs (\ref{Bstruct}). It is tempting therefore to call $C^{d+1}$ a ``space-action''.

\goodbreak

Notice the geometric ``duality'' between the definitions of a NC and of a Carroll structure: while
the first one involves
Newtonian time, $t$, the Carroll structure
involves the ``dual'' or ``Carrollian time'', $s$. Accordingly,
NC structures involve the ``clock'' one-form $\theta$,
while the Carroll structure singles out a vector field $\xi$.
Less trivial and for their physical applications important examples of a Carroll manifolds
can be constructed out of a curved space \cite{DGHunpub}.

The isometry group of the degenerate Carrollian metric $\rg$ is infinite-dimensional since the latter is invariant under the mappings
\beq
x'^A=x^A,
\qquad
s' = s + f(x^1,\ldots,x^d)
\eeq
for any $A=1,\ldots,d$, and arbitrary smooth function $f$. Requiring the preservation of the affine connection, $\nabla$, implies that $f=\const$
Then the automorphisms of the flat Carroll structure~(\ref{flatC}) constitute
the finite-dimensional Carroll group \cite{Leblond, DGH} we  denote simply by $\Carr(d+1)$.
The latter is represented by the matrices
\begin{equation}
a =
\left(
\begin{array}{ccc}
R&0&\bc\\
-\bb^T{}R&1&f\\
0&0&1
\end{array}
\right)\in\Carr(d+1),
\label{Carr}
\end{equation}
where $R\in\rO(d)$, $\bb,\bc\in\IR^d$, and $f\in \mathbb{R}$. Here
the superscript ``$T$'' denotes transposition.

The action of $a\in\Carr(d+1)$ on flat Carroll space-``time'', $C\cong\IR^{d+1}$, thus reads
\begin{equation}
a_C:\left(\begin{matrix}
\bx\\s
\end{matrix}
\right)
\mapsto
\left(\begin{matrix}
R\,\bx+\bc\\
s-\bb^TR\,\bx+f\end{matrix}
\right)
\label{CarrAction}
\end{equation}
for all $\bx\in\IR^d$, and $s\in\IR$.

\goodbreak

Again, we notice that the homogeneous Carroll group spanned by the matrices
\begin{equation}
\widetilde{h} =
\left(
\begin{array}{cc}
R&0\\
-\bb^TR&1
\end{array}
\right)\in\trE(d)
\label{homCarr}
\end{equation}
form a group isomorphic to the Euclidean group (\ref{homGal}); the isomorphism $\rE(d)\to\trE(d)$ is plainly given by
\begin{equation}
\widetilde{h}=(h^T)^{-1}.
\label{HomGalIsomHomCarr}
\end{equation}
Let us emphasize that a Carroll boost $\bb\in\bbR^3$ in (\ref{Carr}) has indeed the physical dimension of a velocity, just as in the Galilei case; this arises from the above-mentioned physical dimension of Carrollian ``time'' $s$.

With these preparations, we can claim that
\begin{equation}
\Carr(d+1)\cong\trE(d)\ltimes\IR^{d+1}.
\label{CarrGlobal}
\end{equation}
The \textit{Carroll Lie algebra}, $\carr(d+1)$, is therefore isomorphic to the Lie algebra of the vector fields
\begin{equation}
X=(\omega^A_B\,x^B+
\gamma^A)\frac{\partial}{\partial{}x^A}+(\varphi-\beta_A\,x^A)\,\frac{\partial}{\partial{}s}\, ,
\label{carralg}
\end{equation}
where $\omega\in\soo(d)$, $\bbeta,\bgamma\in\bbR^d$, and $\varphi\in\bbR$. (Note, once more, that the infinitesimal ``Carrollian boosts'' parametrized by $\bbeta$ act on $C^{d+1}$ consistently with (\ref{carrboost})).
We also record, for later use, the matrix
 representation of the Lie algebra (\ref{carralg}), namely
\begin{equation}
Z =
\left(
\begin{array}{ccc}
\bomega&0&\bgamma
\\
-\bbeta^T&0&\varphi\\
0&0&0
\end{array}
\right)\in\carr(d+1).
\label{Z}
\end{equation}

\goodbreak
For completeness, let us mention that the generators of the Carroll Lie algebra (\ref{carralg}), namely
\beq
P_A=\partial_A,\qquad
J_A=\epsilon_{ABC}\,x_B\,\partial_C,
\qquad K_A=-x_A\partial_s,
\qquad
P_0= \partial_s
\eeq
satisfy the following commutation relations
\beq\barraynb{lllllllll}
[J_A,J_B]&=&-\epsilon_{ABC}J_C,
&[J_A,K_B]&=&-\epsilon_{ABC}K_C,
&[K_A,K_B]&=&0,
\\[6pt]
[J_{A},P_{B}]&=&-\epsilon_{ABC}P_C,
&[K_A,P_B]&=&\delta_{AB}P_0,
&[J_A,P_0]&=&0,
\\[6pt]
[K_A,P_0]&=&0,
&[P_A,P_B]&=&0,
&[P_A,P_0]&=&0.
\earraynb
\label{Carrolrel}
\eeq
for all $A,B=1,\ldots,d$.
\goodbreak

\section{Unification: Bargmann, Newton-Cartan, Carroll}\label{Unifchap}

We now ascend to a \emph{relativistic} spacetime --- but one in one dimension higher.

\subsection{Bargmann manifolds}\label{BargmannSection}

Let us recall first that a \emph{Bargmann manifold} is a triple
$(\IB,\BG,\xi)$,
 where $\IB$ (for Bargmann) is a $(d+2)$-dimensional
manifold with $\BG$ a
metric of signature $(d+1,1)$, and the ``vertical'' vector, $\xi$, a nowhere vanishing, complete, null vector, which is parallel-transported by the Levi-Civita
con\-nection,~$\nabla$, of $\BG$ \cite{DBKP,DGH}.

The \textit{flat} Bargmann structure is given, in an adapted coordinate system, by
\begin{equation}
\IB=\bbR^d\times\bbR\times\bbR,
\quad
\BG=\sum_{A,B=1}^d\delta_{AB}\,dx^A\otimes{}dx^B + dt\otimes{}ds+ds\otimes{}dt,
\quad
\xi=\frac{\partial}{\partial s}\,.\;
\label{Bstruct}
\end{equation}
Note that both
$s$ and $t$ are light-cone, i.e., null, coordinates \footnote{Since the metric $G$ has dimension a length squared, and $t$ that of a time, the new coordinate $s$ has therefore that of action per mass.}.

Factoring out flat Bargmann space, $B$, by the ``vertical'' translations generated by $\xi$, the $(d+1)$-dimensional quotient, $N=B/\bbR\xi$, acquires a flat \textit{Newton-Cartan structure} \cite{DBKP,DGH}.

Likewise, the one-parameter family $C_t\subset{}B$ of $(d+1)$-dimensional sections $t=\const$ admit the \textit{same} flat \textit{Carroll structure} (\ref{flatC}) for all $t\in\IR$ \cite{DGH}.

We will see below, in Sections \ref{BtoNC} and~\ref{CarrInBargSection}, how this comes about in full generality.

Let us recall that the $\xi$-preserving isometries of the flat Bargmann structure (\ref{Bstruct}), namely those diffeomorphisms, $a$, such that
\beq
a^*\BG=\BG,
\qquad
a_*\xi=\xi
\label{BargBarg}
\eeq
form the extended Galilei group \cite{LLcoho} also called \emph{Bargmann group} $\Barg(d+1)$ \cite{DBKP,DGH} of dimension $\half(d^2+3d+4)$, i.e., the group of those matrices of the form
\begin{equation}
a =
\left(
\begin{array}{cccc}
R&\bb&0&\bc\\
0&1&0&e\\
-\bb^T{}R&-\half\bb{}^2&1&f\\
0&0&0&1
\end{array}
\right)\in\Barg(d+1,1),
\label{Barg}
\end{equation}
where $R\in\rO(d)$, $\bb,\bc\in\IR^d$, and $e,f\in\IR$. The Bargmann Lie algebra $\barg(d+1)$ is hence isomorphic to the Lie algebra of the  vector fields of $\IB$,
\begin{equation}
X=(\omega^A_B\,x^B+\beta^A\,t+\gamma^A)\frac{\partial}{\partial{}x^A}+\varepsilon\frac{\partial}{\partial{}t}+(\varphi-\beta_A\,x^A)\,\frac{\partial}{\partial{}s}\in\barg(d+1),
\label{carr}
\end{equation}
where $\omega\in\soo(d)$, $\bbeta,\,\bgamma\in\bbR^d$, and $\varepsilon,\varphi\in\bbR$.

\goodbreak

\subsection{Family tree of groups}

Let us now unveil the relationship between the previous three automorphism groups of the flat structures.

$\bullet$ The Bargmann group (\ref{Barg}) is a non-trivial central extension of the Galilei group~(\ref{Gal}): we have the following group homomorphism
\begin{equation}
\pi:\Barg(d+1,1)\to\Gal(d+1)
\qquad
\hbox{where}
\qquad
\pi(A,\bb,\bc,e,f)=(A,\bb,\bc,e).
\label{BtoG}
\end{equation}

$\bullet$ The Carroll group turns out to be the \textit{derived group} (or the group of commutators) of the Bargmann group
that is, $\Carr(d+1)\cong[\Barg(d+1,1),\Barg(d+1,1)]$; we have hence a group homomorphism
\begin{equation}
\iota:\Carr(d+1)\hookrightarrow\Barg(d+1,1)
\qquad
\hbox{where}
\qquad
\iota(A,\bb\,\bc,f)=(A,\bb\,\bc,0,f),
\label{CinB}
\end{equation}
 with, again, the same notation as before.

Note that while our duality (\ref{tsinterchange}) correspond to the isomorphism $\rE(d)\to\trE(d)$ in (\ref{HomGalIsomHomCarr}) for the homogeneous subgroups, the full groups are not isomorphic, since the homogeneous group $\rE(d)$ acts then differently on the translation subgroup.

\subsection
{Newton-Cartan as the base of Bargmann space}\label{BtoNC}

Call indeed $\vartheta=\BG(\xi)$ the $1$-form associated to $\xi$ on the general Bargmann manifold $(\IB,\BG,\xi)$ introduced in Section \ref{Unifchap}.
Being regular, the covariant symmetric $2$-tensor $\BG=G_{ab}(x)\,dx^a\otimes{}dx^b$ thus admits an inverse $\BG^{-1}=G^{ab}(x)\,\partial_a\otimes\partial_b$, where $(\BG^{-1})^{ab}\BG_{bc}=\delta^a_{c}$.
Since~$\xi$ is automatically an infinitesimal isometry of $(B,G)$, we have $L_\xi\,\BG^{-1}=0$. The contravariant symmetric $2$-tensor $\BG^{-1}$ thus projects to $\IN$, the quotient  of $\IB$
by vertical translations generated by $\xi$  as the contravariant tensor field
$\gamma$ of rank $d$.
 Similarly,  $\vartheta=\BG(\xi)$ is the pull-back to $\IB$ of a ``clock'' $1$-form~$\theta$ on the quotient $N$.
It has, finally, been shown that the Levi-Civita connection, $\nabla$, of $\IB$ naturally defines an affine symmetric connection $\nabla^{\IN}$ on $\IN$ that parallel transports the Galilei structure $(\gamma,\theta)$. A Bargmann structure, $(B,G,\xi)$, thus projects onto a NC structure $(\IN,\gamma,\theta,\nabla^{\IN})$. See \cite{DBKP}.

\subsection
{Carroll as a null hyper-surface embedded into Bargmann space}\label{CarrInBargSection}
Consider now, on $\IB$, the $(d+1)$-dimensional distribution defined by
$\ker\vartheta$, which is indeed the orthogonal complement of $\xi$, and  is, again, integrable since $d\vartheta=0$.
(The ``clock'' $1$-form, $\theta$ is locally of the form $\vartheta=dt$.)
Notice that the  ``vertical'' vector field $\xi$  belongs to this foliation, since $\vartheta(\xi)=\BG(\xi,\xi)=0$.
Call
\begin{equation}
\iota:\IC\hookrightarrow{}\IB
\label{Ct}
\end{equation}
the imbedding at $t=0$, say, of a leaf of $\ker\vartheta$.

Let us now show that the imbedding
(\ref{Ct}) endows $\IC$ with a Carroll structure~\footnote{There is, actually, a whole $1$-parameter family of Carroll manifolds, $C_t$, parametrized by the time values $t\in{}K$. We will write $C=C_0$ with the choice of an origin $0\in{}K$. Something similar occurs ``below'', at the Newton-Cartan level: we have a $1$-parameter family of Riemannian (Euclidean, say) $d$-dimensional manifolds in space-time which have  the same value $t$ of the time coordinate inside a Newton-Cartan manifold. So, Carroll plays inside Bargmann the same r\^ole as Euclid does within Newton-Cartan.}.
Indeed, let us endow $\IC$ with the induced symmetric covariant $2$-tensor
$
\rg^{\IC}=\iota^*\BG,
$
which is degenerate and of rank $d$, since $\ker\rg^{\IC}$ is generated by $\xi$.
At last, let us posit
$
\nabla^{\IC}_XY=\nabla_XY
\,
$ for all $X,Y\in\ker\vartheta.
$
It is a trivial matter to check, using $\nabla\vartheta=0$, that $\vartheta(\nabla^{\IC}_XY)=0$, implying that  $\nabla^{\IC}_XY$ belongs to $\ker\vartheta$. Thus $\nabla^{\IC}$ defines an affine symmetric connection on $\IC$, uniquely associated with the  Levi-Civita  connection, $\nabla$, of $(\IB,\BG)$. This connection also satisfies $\nabla^{\IC}\rg^{\IC}=0$, as well as $\nabla^{\IC}\xi=0$. Thus $(\IC,\rg^{\IC},\xi)$ is a Carroll manifold in the sense of Section \ref{CarrollSection}. The flat Bargmann structure (\ref{Bstruct}) readily
yields the standard flat Carroll structure (\ref{flatC}).

In what follows the superscript ``$C$'' will be dropped wherever no confusion can occur.

\section{Galilei and Carroll \textit{versus} Maxwell}\label{GCvM}


Although the very origin of relativity lies in
Maxwell's electrodynamics, non-Einsteinian
limits can nevertheless be considered
\cite{LBLL}. As Galilean electromagnetism is quite well-known, we will only present below some highlights for the sake of comparison with the Carrollian version to be developed in Section \ref{Carelec}.

 \subsection{Galilean electromagnetism}\label{Galelec}

As observed by Le Bellac and L\'evy-Leblond in the early seventies \cite{LBLL},
Maxwell's electro\-magnetism admits  \emph{two different Galilean limits}, namely the ``magnetic type'',
\beqa
\left\{\barraynb{llllllll}
{\rot}\, {\vE}_m+\displaystyle\frac{\partial{\vB}_m}{\partial t}&=& 0\,,&\qquad&
&{\dive}\,{\vB}_m&=&0,\qquad  \,
 \\[8pt]
{\rot}\, {\vB}_m
&=&0 \,,
&\qquad&
&{\dive}\, {\vE}_m &=&  0\,,\quad
\earraynb \right.
&\hbox{(magnetic type)}
 \label{MGalMax}
\eeqa
which has magnetic induction, but where the
displacement current is missing from Amp\`ere's law, and the ``electric type'',
\beqa
\left\{\barraynb{llllllll}
{\rot}\, {\vE}_e&=& 0\,,&\qquad&
&{\dive}\,{\vB}_e&=&0,\qquad  \,
 \\[8pt]
{\rot}\, {\vB}_e-\displaystyle\frac{\partial{\vE}}{\partial t}
&=&0,
&\qquad&
&{\dive}\,{\vE}_e&=&  0\,,\quad
\earraynb\right.
&\hbox{(electric type)}
 \label{EGalMax}
\eeqa
which has displacement current in Amp\`ere's law, but where the magnetic induction term is missing from Faraday's law.
Then Galilean symmetry is proved, in each case, using the appropriate implementation of Galilean boosts, namely
\beqa
\left\{\barraynb{lllll}
\vB_m(\bx,t) &\to& {\vB}_m'(\bx,t)&=&{\vB}_m(\bx-\bb\,t,t),
\\
{\vE}_m(\bx,t) &\to& {\vE}_m'(\bx,t) &=&
{\vE}_m(\bx-\bb\,t,t) - \vb \times{\vB}_m(\bx-\bb\,t,t)
\earraynb\right.
\;
&\hbox{(magnetic implementation)}\qquad
\label{Mimp}
\eeqa
as well as
\beqa
\left\{\barraynb{lllll}
{\vE}_e(\bx,t) &\rightarrow& {\vE}_e'(\bx,t)&=& {\vE}_e(\bx-\bb\,t,t)\,,
\\
{\vB}_e(\bx,t) &\rightarrow& {\vB}_e'(\bx,t)
&=&{\vB}_e(\bx-\bb\,t,t)+\vb\times{\vE}_e(\bx-\bb\,t,t)\,
\earraynb\right.
\;
&\hbox{(electric implementation)}\qquad
\label{Eimp}
\eeqa
for all $\bb\in\bbR^3$.

\subsection{Carrollian electromagnetism }\label{Carelec}

The Carrollian limit of the Maxwell equations can also be considered.
Following Ref. \cite{GiHaYi} we start with
the  vacuum Maxwell equations,
\beq
\left\{
\barraynb{lllllllll}
\ \ \ \ \ {\rot}\, {\vE}&+
\displaystyle\frac{\partial{\vB}}{ \partial t}
&=& 0\,,&\qquad
\qquad&&{\dive}\,{\vB}&=&0,\qquad  \,
 \\[8pt]
-c^2 \,{\rot}\, {\vB}
&+\displaystyle\frac{\partial{\vE}}{\partial t}
&=&0,
&\qquad
&\qquad&{\dive}\, {\vE} &=&  0,\,\qquad
\earraynb
\right.
\label{Maxeq}
\eeq %
where $t$ is relativistic time.

\goodbreak

Letting here $c\to\infty$ would then yield the magnetic Galilean limit  (\ref{MGalMax}) with $\vE$ and $\vB$ unchanged.
Redefining the fields instead
$
\vB\to \vB_e=c\,\vB,
\;
{\vE}\to \vE_e=\vE/c
$
and letting $c\to\infty$ would provide us with the electric limit,
(\ref{EGalMax}).

Let us now investigate the \textit{Carrollian limit} of Maxwell's equations (\ref{Maxeq}) by considering $s$ in (\ref{Carrtime}) as ``time'', instead of $t$.
 \emph{After a re-definition of the electro-magnetic field},
\beq
\widetilde{\vE}=\vE,
\qquad
\widetilde{\vB}=(c\,C)\,\vB,
\eeq
the Maxwell equations are re-written as
\beq
\left\{\barraynb{lllllllll}
\ \ \ {\rot}\, \widetilde{\vE}&+\;\;\;\;\;\;
\displaystyle\frac{\partial{\widetilde{\vB}}}{\partial s}
&=& 0,\,&\qquad
\qquad&&{\dive}\,\widetilde{\vB}&=&0,\qquad  \,
 \\[8pt]
-\;{\rot}\, \widetilde{\vB}
&+\;C^2 \,\,\displaystyle\frac{\partial\widetilde{\vE}}{\partial s}
&=&0,
&\qquad
&\qquad&{\dive}\, \widetilde{\vE}&=&  0.\,\qquad
\earraynb \right.
 \label{MaxCeqs}
\eeq
Hence
\beq
\left[\bigtriangleup\;-\;C^2\left(\frac{\p}{\p{}s}\right)^2\right]
\barray{c}
\widetilde{\vE}
\\
\widetilde{\vB}
\earray=0,
\eeq
which allows us to interpret $C^{-1}$ as the \emph{propagation velocity of electromagnetic waves measured in ``time'' $s$}, i.e., the \emph{speed of light} with respect to $s$. Let us observe that the physical dimension of the constant $C$ is $[C]=[s]/L=LT^{-1}$,
i.e., a velocity, as it should be.  The two ``time" coordinates are hence proportional, with scaling factor the quotient of light speeds in both theories.

\subsubsection{Electric-like contraction}

Moreover, taking the Carrollian limit
$C\uparrow\infty$ (with $\widetilde{\vE}$ and $\widetilde{\vB}$ fixed) switches off the Amp\`ere term
${\rot}\,\widetilde{\vB}$,
providing us with the equations of ``Carrollian electromagnetism of the electric type'',
\beq\left\{\barraynb{lllllll}
{\rot}\,\widetilde{\vE}_e+\displaystyle\frac{\partial\widetilde{\vB}_e}{\partial s}\,&=& 0,&\qquad
&{\dive}\,\widetilde{\vB}_e&=&0
 \\[8pt]
\hfill\displaystyle\frac{\partial\widetilde{\vE}_e}{\partial s}
&=&0\,,
&\qquad
&{\dive}\,\widetilde{\vE}_e&=&  0\
\earraynb\right.
\qquad\hbox{(electric type)}
 \label{ECMax}
\eeq %
where $\widetilde{\vE}_e=\widetilde{\vE}$ and $\widetilde{\vB}_e=\widetilde{\vB}$.

This theory is Carroll-invariant, as expected.
 Carrollian boosts (\ref{carrboost}), implemented as
\beqa
\left\{\barraynb{lllll}
\widetilde{\vE}_e(\bx,s) &\rightarrow& \widetilde{\vE}_e'(\bx,s)&=& \widetilde{\vE}_e(\bx,\,s-\bb\cdot \bx)\,
\\
\widetilde{\vB}_e(\bx,s) &\rightarrow& \widetilde{\vB}_e'(\bx,t)
&=&\widetilde{\vB}_e(\bx,\,s-\bb\cdot \bx)+\vb\times\widetilde{\vE}_e(\bx,\,s-\bb\cdot \bx)\,
\earraynb\right.
\qquad
&
\label{CarEimp}
\eeqa
are readily shown to leave (\ref{ECMax})  invariant.
Let us observe that (\ref{CarEimp}) is in fact
an \emph{electric-type} implementation (\ref{Eimp}),
 as anticipated by our labeling -- and that despite the presence of the Faraday term in (\ref{ECMax}).

The Carrollian Maxwell equations (\ref{ECMax})
can be derived from an action principle as follows. The usual relativistic action is
\beq
S = \int\half\bigl({\vE}^2-c^2{\vB}^2\bigr)\,dt\,d^3\vx
=
\big(cC\big)^{-1}\int\half\bigl(\widetilde{\vE}^2-\frac1{C^2}\widetilde{\vB}^2\bigr)\,ds\,d^3\vx,
\label{usaction}
\eeq
with
$
\widetilde{\vB} = {\rot}\, \widetilde{\bA}\,,
\;
\widetilde{\vE} =-{\grad}\widetilde\phi-{\p\widetilde{\bA}}/{\partial s}\,.
$
Dropping the pre-factor $\big(cC\big)^{-1}$ and taking the limit $C \uparrow \infty$ provides us with the action
\beq
S_e =\int\half\,\widetilde{\vE}_e^2\, dsd^3\vx,
\label{eaction}
\eeq
whose variation gives the second line of (\ref{ECMax}), while the first line follows from our using the potentials.

This result also confirms that the system (\ref{ECMax}) is indeed an electric-type theory.

\subsubsection{Magnetic-like contraction}

A \emph{magnetic-type} Carroll-invariant version of the Maxwell equations can also be found, though. It is an easy matter to prove indeed that the system
\beq
\left\{\barraynb{lllllll}
{\rot}\, \widetilde{\vB}_m-\displaystyle\frac{\p\widetilde{\vE}_m}{\partial s}&=&0\,,&\qquad
&{\dive}\,\widetilde{\vE}_m&=&0,
 \\[8pt]
\displaystyle
\hfill\frac{\partial\widetilde{\vB}_m}{\partial s}
&=&0,
&\qquad
&{\dive}\,\widetilde{\vB}_m
&=&
0,
\earraynb
\right.\,
\qquad
\hbox{(magnetic type)}
\label{MCMax}
\eeq %
is also invariant, provided  Carroll-boosts  act by the {magnetic-type implementation}
\beqa
\left\{\barraynb{lllll}
\widetilde{\vE}_m(\bx,s) &\rightarrow& \widetilde{\vE}_m'(\bx,s)&=& \widetilde{\vE}_m(\bx,\,s-\bb\cdot \bx)
-\vb\times\widetilde{\vE}_m(\bx,\,s-\bb\cdot \bx)\,,
\\
\widetilde{\vB}_m(\bx,s) &\rightarrow& \widetilde{\vB}_m'(\bx,t)
&=&\widetilde{\vB}_m(\bx,\,s-\bb\cdot \bx),
\earraynb\right.
\qquad
&
\label{CarMimp}
\eeqa
cf. (\ref{Mimp}).

\goodbreak

Remember that the  relativistic Maxwell equations, (\ref{Maxeq}) and (\ref{MaxCeqs}), respectively, are invariant under \emph{electric-magnetic duality transformation},
\beq
\left\{\barraynb{lll}
{\vE}&\to&c^2{\vB}
\\
{\vB}&\to&-{\vE}
\earraynb\right.
\qquad\hbox{i.e.}\qquad
\left\{\barraynb{lll}
\widetilde{\vE}&\to&\widetilde{\vB}
\\
\widetilde{\vB}&\to&-C^2\widetilde{\vE}
\earraynb\right.
\label{emduality}
\eeq
Taking either the Galilean  and resp. the Carrollian limit, $c\uparrow\infty$
resp.  $C\uparrow\infty$, breaks
this symmetry:
\beq
\left\{\barraynb{lll}
{\vE}_m&\to& {\vB}_e
\\
{\vB}_m&\to&-{\vE}_e
\earraynb\right.
\qquad\hbox{and}\qquad
\left\{\barraynb{lll}
\widetilde{\vE}_m&\to&\widetilde{\vB}_e
\\
\widetilde{\vB}_m&\to&-\widetilde{\vE}_e
\earraynb\right.
\label{NEemduality}
\eeq
intertwine instead, in both the Galilean and  Carrollian cases, the ``magnetic type''  equations (\ref{MGalMax}) and
(\ref{MCMax}) with the
 ``electric type'' ones, (\ref{EGalMax}) and (\ref{ECMax}), respectively.
The implementations (\ref{Mimp}) and (\ref{Eimp}) as well as (\ref{CarEimp}) and (\ref{CarMimp}), respectively, are also interchanged.

The precise structure will be clarified  in Section \ref{GeoCarr}  below.

\section{Geometric formulation and symmetries of Carroll electromagnetism}\label{GeoCarr}

Our Carrollian theories can also be presented in a geometric framework.
To motivate what follows, let us first
 recall some aspects of the full Maxwell theory.

\goodbreak

The source-free Maxwell equations on a $(d+1)$-dimensional space-time, $(M,\rg)$, with Lorentz signature involve both covariant and contravariant objects,
\beqa
dF=0
\label{Max1}
\\
\Div_\rg(F^\sharp)=0
\label{Max2}
\eeqa
where the \emph{$2$-form} $F=\half{}F_{ab}\,dx^a\wedge{}dx^b$ is the electromagnetic field, and
$F^\sharp$ is the \emph{bi-vector}
\footnote{Considering the Maxwell electromagnetic bi-vector is purely formal in the Lorentzian framework; this will however prove crucial later on, in the Carrollian setting.},
\begin{equation}
F^\sharp=\half{}F_\sharp^{ab}\,\partial_a\wedge\partial_b,
\quad\hbox{where}\quad
F_\sharp^{ab}
=
\rg^{ac}\rg^{bd}\,F_{cd},
\label{FsharpMaxwell}
\end{equation}
and $\Div_\rg$ is the covariant divergence,
\begin{equation}
\nabla_aF_\sharp^{ab}=\partial_a{}F_\sharp^{ab}+\Gamma^a_{ac}F_\sharp^{cb}.
\label{Max2Bis}
\end{equation}
for all $b=0,\ldots n=d$, where $\nabla$ stands for the Levi-Civita connection.

\goodbreak

Writing locally,
$
F=E_1 dx^1\wedge{}dt+\cdots+B_1 dx^2\wedge{}dx^3+\cdots
$
where $t=x^0$, we know that Eqns (\ref{Max1}) and (\ref{Max2}) reduce to the ordinary free Maxwell equations in Minkowski spacetime, $\bbR^{3,1}$.

In  usual (relativistic)  Maxwell  theory $2$-forms and bi-vectors are equivalent, since by the ``musical isomorphism''
one can pass from one to the other by simple ``index gymnastics'', using the Lorentz metric, cf.  (\ref{FsharpMaxwell}).
This is \emph{not} the case in non-Einsteinian
physics, though, where, owing to the degeneracy of the
(Galilean or Carrollian)  ``metric'',
covariant and contravariant vectors can not be
converted freely into each other.

Considering then a $(d+1)$-dimensional Carroll space-time manifold $(C,\rg,\xi,\nabla)$, we will try and reproduce below, in this new geometrical framework, what K\"unzle did to formulate intrinsically the two Le~Bellac-L\'evy-Leblond versions of Galilean electromagnetism for Newton-Cartan structures~\cite{Kun2}.

\subsection{Contravariant Carroll theory}\label{CMcontra}

We start  with the electromagnetic field viewed as a \textit{bi-vector}
\begin{equation}
F_m=\half{}F^{ab}_m\,\partial_a\wedge\partial_b,
\label{FContra}
\end{equation}
where the subscript``$m$'' stands for ``magnetic'' --- as will be justified below ---, and use the Carroll ``metric'', $\rg$, of Carroll spacetime $(C,\rg,\xi,\nabla)$ to  define the associated $2$-form to lower indices,
\begin{equation}
F^\flat=\half(F^\flat_m)_{ab}\,dx^a\wedge{}dx^b
\qquad
\hbox{where}
\qquad
(F^\flat_m)_{ab}=\rg_{ac}\,\rg_{bd}\,F^{cd}_m.
\label{Fflat}
\end{equation}
Note that the ``lowering operator'', ``$\flat$'', such that $\flat(F_m)=F^\flat_m$
converts contravariant objects, e.g., bi-vectors,  into co\-variant tensors, e.g., $2$-forms.

Then, to mimic the \textit{homogeneous Maxwell equations},  we require that $F^\flat_m$ be closed, viz.,
\begin{equation}
dF^\flat_m=0
\qquad\Longleftrightarrow\qquad
\partial_{[a}(F^\flat_m)_{bc]}=0,
\label{CarContra1}
\end{equation}
for all $a,b,c=0,\ldots,d$.

Likewise, wanting to reproduce  \textit{``inhomogeneous'' Maxwell equations}, we posit
\begin{equation}
\Div (F_m)=0
\qquad\Longleftrightarrow\qquad
\nabla_a{}F^{ab}_m=0
\label{CarContra2}
\end{equation}
for all $b=0,\ldots,d$.
The system (\ref{CarContra1}) -- (\ref{CarContra2}) of PDE for the bi-vector $F_m$ will constitute the \textit{``contravariant-type'' Carroll-Maxwell equations} in vacuum.

In the $(3+1)$dimensional flat Carroll space-time (\ref{flatC}), putting
\begin{equation}
F_m=E^A\,\partial_A\wedge\partial_s
+\half\epsilon^{ABC}B_C\,\partial_A\wedge\partial_B,
\label{Fm}
\end{equation}
where $\epsilon^{ABC}$ is the standard Levi-Civita symbol, we find that $F^\flat_m$ is purely magnetic, viz.,
\begin{equation}
F^\flat_m=
\half\epsilon_{ABC}B^C\,dx^A\wedge{}dx^B,
\label{FflatBis}
\end{equation}
where $B_C=B^C$ for all $C=1,2,3$. So, the system (\ref{CarContra1}) -- (\ref{CarContra2})  for the \textit{contravariant electro\-magnetic field} $F$ become precisely the ``magnetic-type'' system (\ref{MCMax}), with a mere change of notation: $\vE\to\tbE_m$, and $\vB\to\tbB_m$.

We now show in general terms that the \textit{Carroll group}, $\Carr(C,\rg,\xi,\nabla)$,
 is actually a group of symmetries of the contravariant-type Carroll-Maxwell equations. We  confine considerations to infinitesimal symmetries of the system (\ref{CarContra1}) -- (\ref{CarContra2}), namely to those vector fields~$X$ of $C$ such~that
\begin{eqnarray}
\label{LXdContra}
L_Xd\circ\flat\,F_m&=&d\circ\flat\,L_XF_m\\[6pt]
\label{LXDIvContra}
L_X\Div{}F_m&=&\Div{}L_XF_m
\end{eqnarray}
for all bi-vectors $F_m$ solutions of Eqs (\ref{CarContra1}) and (\ref{CarContra2}). Equation (\ref{LXdContra}) holds identically since $L_X(F^\flat_m)=(L_XF_m)^\flat$ for any Carroll generator~$X$ in view of~(\ref{carrC}). It simply remains to prove that Equation (\ref{LXDIvContra}) holds for any $X\in\carr(C,\rg,\xi,\nabla)$. Indeed, straightforward calculation shows that
\begin{equation}
\left(\left[L_X,\Div\right]F_m\right)^b=L_X\Gamma^a_{ac}\,F^{cb}_m
\label{CommLXDiv}
\end{equation}
for all $b=1,\ldots,n$. At last, Carroll automorphisms being affine, $L_X\nabla=0$, Eq.~(\ref{LXDIvContra}) is verified.

\subsection{Covariant Carroll theory}\label{CovCarrSection}

The covariant theory admits a slightly
 more subtle formulation.
Here we start with the \emph{covariant} electromagnetic \emph{2-form},
\begin{equation}
F_e=\half{}F_{ab}\,dx^a\wedge{}dx^b,
\label{FCov}
\end{equation}
where the subscript``$e$'' means now ``electric''.

\goodbreak

To produce a \emph{bi-vector} designed to enter the ``inhomogeneous'' Maxwell-Carroll field equations, we resort to the only contravariant object at hand, namely to the vector field $\xi$. Therefore we
consider  the $1$-form $E^\flat=-F_e(\xi)$ obtained by contracting $F_e$ with $\xi$,
and then converting it to a vector by
using the Carroll metric, $E^\sharp=\rg^{-1}(E^\flat)$. Then putting
\beq
F^\sharp_e=E^\sharp\wedge\xi
\label{Fsharp}
\eeq
provides us with a well-defined \emph{bi-vector}.
Let us work out a coordinate expression for $F^\sharp_e$
via some ``generalized inverse'', $\rg_\varphi=\rg_\varphi^{ab}\,\partial_a\otimes\partial_b$, of the degenerate ``metric'' $\rg$ on Carroll space-time~$(\IC,\rg,\xi)$. This twice-symmetric contravariant tensor field $\rg_\varphi$ is defined in a unique fashion by the equations
$\rg_\varphi^{ak}\,\rg_{kb}=\delta^a_b-\xi^a\varphi_b$, where~$\varphi$ is a $1$-form such that $\rg_\varphi^{ab}\varphi_b=0$ for all $b=0,\ldots,n=d$ (implying $\varphi_a\xi^a=1$). Having chosen such a $\rg_\varphi$,  put $E_\sharp^a=\rg_\varphi^{ak}E_k$ where $E_k=-F_{ak}^e\xi^a$ as above. The ``electric'' bi-vector $F^\sharp_e=\sharp(F_e)$ in (\ref{Fsharp}) takes, hence, the local form
\begin{equation}
F^\sharp_e=\half({F^\sharp_e})^{ab}\,\p_a\wedge\p_b
\qquad
\hbox{where}
\qquad
({F^\sharp_e})^{ab}=2\,\rg_\varphi^{k[a}\,\xi^{b]}\,F_{k\ell}\,\xi^\ell,
\label{FsharpLocal}
\end{equation}
which, moreover, turns out to be independent of the $1$-form $\varphi$ \footnote{Note that $\rg_\varphi^{ab}=\rg_\psi^{ab}+2\xi^{(a}\eta^{b)}$ where the vector $\eta$ depends explicitly on the $1$-forms $\varphi$ and $\psi$.}. (Compare to the Maxwel\-lian expression (\ref{FsharpMaxwell}).)
Note that the ``raising operator''  ``$\sharp$''
converts covariant objects ($2$-forms) into
contravariant ones, namely bi-vectors.

At last, we  chose $F^\sharp_e$ as the contravariant counterpart of  the covariant electromagnetic field $F_e$ in (\ref{FCov}) and posit the following field equations, namely
\beq
\left\{
\barraynb{lll}
dF_e=0,\\[4pt]
\Div(F^\sharp_e)=0
\qquad\Longleftrightarrow\qquad
\nabla_a({F^\sharp_e})^{ab}=0,
\earraynb
\right.
\label{CarCov}
\eeq
for all $b=0,\ldots,d$.

The system (\ref{CarCov}) -- (\ref{CarContra2}) of PDE for the $2$-form $F_e$ will constitute the \textit{``covavariant-type'' Carroll-Maxwell equations} in vacuum.

If we write, locally, in the $(3+1)$-dimensional flat Carroll spacetime (\ref{flatC}),
\begin{equation}
F_e=E_A\,dx^A\wedge{}ds+
\half\epsilon_{ABC}\,B^C\,dx^A\wedge{}dx^B,
\label{FCovBis}
\end{equation}
so that $E^\flat=E_A dx^A$, then the associated bi-vector (\ref{Fsharp}) reads now
\begin{equation}
F^\sharp_e=E^A\,\partial_A\wedge\partial_s,
\label{FsharpBis}
\end{equation}
and is ``purely electric'' (here, $E^A=E_A$ for all $A=1,2,3$).
The covariant Maxwell-Carroll equations (\ref{CarCov}) readily become the ``electric-type'' equations (\ref{ECMax}) once we rename $\vE\to\tbE_e$, and $\vB\to\tbB_e$.

The symmetries of the covariant-type Carroll-Maxwell equations (\ref{CarCov}) can again be studied in geometric terms, much in the same way as in Section \ref{CMcontra}.
Those consist in the vector fields $X$ of~$C$ that preserve the equations (\ref{CarCov}), namely such that the following commutators of differential operators vanish, namely
\begin{eqnarray}
\label{LXdCov}
L_XdF_e&=&dL_XF_e\\[6pt]
\label{LXDIvCov}
L_X\Div\circ\sharp\,F_e&=&\Div\circ\sharp\,L_XF_e
\end{eqnarray}
for all $2$-forms $F_e$ solutions of Eqs (\ref{CarCov}). Just as before, Equation (\ref{LXdCov}) is identically verified. We can again prove that
\begin{equation}
([L_X,\Div\circ\sharp]F_e)^b=L_X\Gamma^a_{ac}(F^\sharp_e)^{cb}
\label{LXDivSharpF}
\end{equation}
for all $b=0,\ldots,d$, and for all $X\in\carr(C,\rg,\xi,\nabla)$. The fact that Carroll transformations are affine entails that Eq. (\ref{LXDIvCov}) is verified by any infinitesimal Carroll automorphism $X$.

We will elsewhere prove that Carroll electromagnetisms admit, in fact a larger, infinite-dimensional, Lie algebra of symmetries \cite{DGHunpub}.

\subsection{Carroll electromagnetisms versus Maxwell theory on Bargmann spaces}

Let us show how the two Carroll electromagnetisms actually \textit{stem from  Maxwell field theory} on Bargmann manifolds introduced in Section \ref{BargmannSection}. \footnote{Much in the same manner, Galilean electromagnetisms arise from plain Maxwell theory on Bargmann spaces. See \cite{DGH} for a detailed account.}

To prove this, let us start with the Maxwell equations on a $(d+1,1)$-dimensional Bargmann manifold $(\IB,G,\xi)$, namely
\beqa
d\cF=0
\label{Max1bis}
\\
\Div_G(\cF^\sharp)=0
\label{Max2bis}
\eeqa
where $\cF$ is a $2$-form on $B$ (see Eqs (\ref{Max1}) and (\ref{Max2})). The Carroll manifold we are dealing with will be given, as in Section (\ref{Ct}), by the embedding $\iota:\IC\hookrightarrow{}\IB$, defined by $t=\const$, say.

\subsubsection{Electric-like case}\label{electricSubSubSection}

The induced $2$-form
\begin{equation}
F_e=\iota^*\cF
\label{FeBis}
\end{equation}
of $\IC$ is clearly closed in view of (\ref{Max1bis}),
$
dF_e=0;
$
this corresponds to the first equation in (\ref{CarCov}).

Consider now the bi-vector $\cF^\sharp$ of $\IB$ defined by $\cF^{\mu\nu}_\sharp=G^{\mu\alpha}G^{\nu\beta}\cF_{\alpha\beta}$ for all $\mu,\nu=0,\ldots,d+1$. Then, the restriction
\begin{equation}
F^\sharp_e=\cF^\sharp\vert\IC
\qquad
\hbox{with}
\qquad
\cF^\sharp(G(\xi))=0,
\label{FesaharpBis}
\end{equation}
of $\cF^\sharp$ to $\IC$ reads clearly $(F^\sharp_e)^{ab}=\rg_\varphi^{ac}\rg_\varphi^{cd}(\cF_{cd}\vert\IC)$ for some $1$-form $\varphi$ of~$\IC$ such that $\varphi(\xi)=1$ (see Section \ref{CovCarrSection}). The second equation in (\ref{FesaharpBis}) is mandatory to duly restrict, e.g., from $10$ to $6$ if $d=3$, the number of components of $\cF^\sharp$. We hence have $\cF_\sharp^{\mu\nu}\xi_\nu=0$ at each point of the Carroll manifold $\IC$, which implies that $F^\sharp_e$ is a well-defined bi-vector of~$\IC$. Now $F^\sharp_e(\varphi)=0$ entails that $F^\sharp_e$ has rank~$\leq2$, and hence $F^\sharp_e=E^\sharp\wedge\xi$ as in (\ref{Fsharp}).
At last, the Carroll connection on~$\IC$ being induced from the Levi-Civita connection of $(\IB,G)$, we find that $\Div(F^\sharp_e)=\Div_G(\cF^\sharp)\vert\IC$; thanks to (\ref{Max2bis}) we end up with
$
\Div(F^\sharp_e)=0,
$
i.e., with the second equation in (\ref{CarCov}) governing \textit{electric-like} Carroll electromagnetism.

\subsubsection{Magnetic-like case}\label{magneticSubSubSection}

Start with the electromagnetic bi-vector $\hcF$ of our Bargmann manifold $B$, whose components read $\hcF^{\mu\nu}=G^{\mu\alpha}G^{\nu\beta}\cF_{\alpha\beta}$ for all $\mu,\nu=0,\ldots,d+1$. \footnote{Obviously $\hcF=\cF^\sharp$, but we want another notation for this bi-vector to avoid confusion and clutter.}

This twice-contravariant tensor will define a well-behaved bi-vector $F_m$ of our Carroll sub\-manifold~$C$ of $\IB$, if we put
\begin{equation}
F_m=\hcF\vert\IC
\qquad
\hbox{with}
\qquad
\hcF(G(\xi))=0,
\label{FmBis}
\end{equation}
which again means that we consistently restrict the number of components of $\hcF$ by imposing the constraints $\hcF^{\mu\nu}\xi_\nu=0$.
Using the above arguments, we readily conclude that $\Div(F_m)=\Div_g(\hcF)\vert\IC$, so that $\Div(F_m)=0$, in full accordance with Eq. (\ref{CarContra2}).

Consider then $2$-form $\hcF^\flat$ associated with the above bi-vector $\hcF$, viz., $\hcF^\flat=G_{\mu\alpha}G_{\nu\beta}\hcF^{\alpha\beta}$, where again $\hcF^{\mu\nu}\xi_\nu=0$, for all $\mu,\nu=0,\ldots,d+1$. Now, we have seen that the induced Bargmann metric on $\IC$ is precisely the Carroll metric; this entails that the $2$-form
\begin{equation}
F_m=\iota^*\hcF^\flat
\label{FmflatBis}
\end{equation}
is, thanks to (\ref{Max1bis}), actually closed, $dF_m=0$, confirming that $F_m$ as defined by (\ref{FmBis}) solves indeed Eq. (\ref{CarContra1}). This ends the proof that the magnetic-like Carroll field equations are deduced from the Maxwell equations (in their contravariant form) on Bargmann ``space-time-action''.


We notice, at last, that the Carroll electric/magnetic duality (\ref{NEemduality}) is plainly given by the correspondence
\begin{equation}
\ast:F_m\to{}F_e
\label{EMCarrollDuality}
\end{equation}
spelled out in the preceding sections.

\section{Non-Einsteinian electrodynamics in a medium}

Returning to a down-to-earth approach, let us remember that, in a medium endowed with electric charge density $\rho$, and current density $\vj$, the  Maxwell equations are written
\beq\left\{\barraynb{llllll}
\vnabla\times {\vE}+\displaystyle\frac{\partial{\vB}}{\partial t}
&=&0,
&\vnabla
\cdot{\vB}&=&0,
\\[12pt]
\vnabla\times{\vH}-\displaystyle\frac{\partial {\vD}}{\partial t}&=&{\vj},%
&\vnabla\cdot {\vD}&=&\rho,
\earraynb\right.
\label{medMeq}
\end{equation}
where ${\vE}$ is the electric field, ${\vD}$  the electric
displacement, ${\vB}$  the magnetic induction, and
${\vH}$ is the magnetic field.
 This system of 8 equations involves $12$ fields,
and additional constraints called \emph{constitutive relations} should therefore be imposed. The standard choice is
\beq
\vD=\epsilon\, \vE,
\qquad
\vB=\mu\,\vH.
\label{usconsteq}
\eeq
This completes (\ref{medMeq}), whose Lorentz invariance can then be proven as it is known from textbooks.
In the vacuum,  $\epsilon=\epsilon_0$ and $\mu=\mu_0$ are constants such that
$\epsilon_0\mu_0=c^{-2}$.

Goldin and Shtelen \cite{GoSh} pointed out, however, that (\ref{usconsteq}), although dictated by physical arguments, in \emph{not}
the only choice which is consistent with Lorentz symmetry. In fact, implementing Lorentz boosts in the usual way and choosing
\beq
{\vD}=\alpha{\vB}+\frac 1{c^2}\beta{\vE},
\qquad
\vH=\beta{\vB}-\alpha{\vE},
\eeq
yields a Lorentz-invariant system,
where $\alpha$ and $\beta$ are arbitrary scalar functions of the first two, namely $I_1$, $I_2$, of the Lorentz invariants,
\beq\barraynb{lllllllll}
I_1&=&{\vB}^2-\displaystyle\frac{1}{c^2}
{\vE}^2,\;\;\;
&I_2&=&{\vB}\cdot{\vE},\;\;\;\;
&I_3&=&{\vD}^2-\displaystyle\frac{1}{c^2}
{\vH}^2,
\\[6pt]
I_4&=&{\vH}\cdot{\vD},\;\;\;
&I_5&=&{\vB}\cdot{\vH}-{\vE}\cdot {\vD},
\;\;\;\;
&I_6&=&{\vB}\cdot{\vD}
+\displaystyle\frac{1}{c^2}{\vE}\cdot{\vH}.
\earraynb
\label{Lorentzinv}
\eeq
The usual choice corresponds plainly to
$\alpha=0$ and $\beta={\mu_0}^{-1}$.

Can the system be made also Galilei-invariant?
The question sounds paradoxical, since relativistic physics has its very roots in the Maxwell equations.  However, as pointed out by Le Bellac and L\'evy-Leblond already, the obstruction against Galilean invariance comes entirely from the \emph{constitutive relations}.
Goldin and Shtelen \cite{GoSh} argue, moreover,
that an appropriate (although unconventional) choice of the latter can make the combined
system \emph{Galilei invariant}, while leaving the
Maxwell (\ref{medMeq})  \emph{unchanged}!
Let us outline how this comes about.

Let us hence consider an ordinary Galilei boost, (\ref{Galboost}). Then a straightforward calculation shows that the magnetic-type implementation (\ref{Mimp}) on $\vE$ and $\vB$
extended to the fields $\vD$ and $\vH$, namely,
\beq\barraynb{llllll}
{\vE}^{\prime }&=&{\vE}-\vb\times{\vB},\;\;\;&{\vB}^{\prime }&=&\vB
\\[2pt]
{\vH}^{\prime}&=&{\vH}+\vb\times {\vD},\;\;\;\;&{\vD}^{\prime}&=&{\vD},
\\[2pt]
{\vj}^{\prime }&=&{\vj}+\rho\vb,\;\;\;\;
&\rho^{\prime }&=&\rho.
\earraynb
\label{medGimp}
\eeq
leaves the system (\ref{medMeq}) invariant.

Then Goldin and Shtelen proceed to prove that
the constitutive equations can also be made consistent with Galilei transformations
${\vx}'=\vx+\vb t$ \cite{GoSh}. Let us show how.

First, one checks that implementing
Galilei transformations on the fields according
to (\ref{medGimp})
yields the Galilean invariants
\beq\barraynb{lllllllll}
I_1&=&{\vB}^2,\;\;\;\;
&I_2&=&{\vB}\cdot {\vE},\;\;\;\;
&I_3&=&{\vD}^2,
\\[6pt]
I_4&=&{\vH}\cdot {\vD},\;\;\;\;
&I_5&=&{\vB}\cdot{\vH}-{\vE}\cdot {\vD},\;\;\;\;\;
&I_6&=&{\vB}\cdot{\vD}.
\label{galinvar}
\earraynb
\eeq
Then a direct calculation shows that the unconventional constitutive relations
\begin{equation}
{\vD}=\hat{\alpha}\,{\vB},\;\;\;\;\;
\vH=\hat{\beta}\,{\vB}-\hat{\alpha}\,{\vE}
\label{Galconstrel}
\end{equation}
where $\hat{\alpha}$ and $\hat{\beta}$ are arbitrary functions of the Galilei invariants in (\ref{galinvar})
makes the combined system (\ref{medMeq})-(\ref{Galconstrel}) Galilei-invariant.

Let us insist that here one works with \emph{unmodified} Maxwell equations; the symmetry  comes entirely from the appropriate choice of the constitutive relations.
The new constitutive relations (\ref{Galconstrel})
are plainly inconsistent with the usual
choice (\ref{usconsteq}) as they should: the
latter are indeed Lorentz, and not Galilei, invariant.

Goldin and Shtelen argue that the Galilean limit they consider could be applied to describe light propagation which, in certain media, can be as slow as 17 m/s \cite{Nature}.

We partly disagree with them: the mentioned
velocity of light is so incredibly low that it is rather the \textit{Carrollian approximation},
\beq
c\downarrow0
\eeq
which would appear more appropriate.
Can we make the system Carroll-invariant by a suitable choice of constituent relations~?
The answer is \emph{positive} as we now show.
Consider indeed the Carrollian version of the
electric-type implementation, (\ref{Eimp}), viz.
\beq\barraynb{llllll}
{\vE}^{\prime }&=&{\vE},\;\;\;
&{\vB}^{\prime}&=&\vB+\vb\times {\vE}
\\[2pt]
{\vH}^{\prime}&=&{\vH},\;\;\;\;
&{\vD}^{\prime}&=&{\vD}-\vb\times{\vH},
\\[2pt]
{\vj}^{\prime }&=&{\vj},\;\;\;\;
&\rho^{\prime }&=&\rho-\vb\cdot{\vj},
\earraynb
\label{medCimp}
\eeq
Then a straightforward calculation shows, that the
Maxwell system (\ref{medMeq}) is left invariant.
\footnote{
One can wonder if a magnetic-type implementation cf. (\ref{Mimp}), does exist in this case also. The answer is \emph{no},
since the first line of the extended system
(\ref{medMeq}) is only consistent with
(\ref{Eimp}), but not with (\ref{Mimp}).}

The next step is to derive the \emph{Carrollian invariants}
\beq\barraynb{lllllllll}
I_1&=&{\vE}^2,\;\;\;\;
&I_2&=&{\vB}\cdot {\vE},\;\;\;\;
&I_3&=&{\vH}^2,
\\[6pt]
I_4&=&{\vH}\cdot {\vD},\;\;\;\;
&I_5&=&{\vB}\cdot {\vH}-{\vE}\cdot {\vD},\;\;\;\;\;
&I_6&=&{\vE}\cdot{\vH}.
\earraynb
\label{Carrolinvar}
\end{equation}
Then, searching for constitutive relations of
the form
$
{\vD}=\alpha{\vB}+\beta{\vE},\;{\vH}=\gamma {\vB}+\delta
{\vE},
$
a direct calculation yields the coefficients,  $\gamma=0,\delta=-\alpha$, providing us with
the general \emph{Carrollian constitutive equations},
\begin{equation}
{\vD}=\alpha\,{\vB}+\beta\,{\vE},\;\;\;\;
{\vH}=-\alpha\,{\vE},
\end{equation}
where $\alpha=\alpha\left( I_1,I_2\right)$ and $\beta=\beta\left(I_1,I_2\right)$ are arbitrary
function of the Carrollian field invariants $I_1$ and~$I_2$ in (\ref{Carrolinvar}).

\subsection{Pre-metric electrodynamics  and  the
Goldin -- Shtelen approach }

In the absence of sources Maxwell's equations
\begin{eqnarray}
\vnabla \times {\vE}+ \frac{\partial{\vB}}{\partial t}&=&0,\;\;\;\;
\vnabla \cdot{\vB}=0\,,  \nn\\
\vnabla\times{\vH}-\displaystyle\frac{\partial {\vD}}{\partial t}&=&0%
,\;\;\;\; \vnabla\cdot {\vD}=0,
\label{mediumMeq}
\end{eqnarray}
may be written as
\begin{eqnarray}
dF=0,
\qquad
F= \half F_{ab} dx^a \wedge dx^b&=-& E_A\, dt \wedge dx^A + \half \epsilon_{ABC}\,B^C dx^A \wedge dx^B,
\nn
\\[6pt]
dH=0,
\qquad
H= \half H_{ab} dx^a \wedge dx^b
&=& H_A\, dt\wedge dx^A +\half\epsilon_{ABC}\,D^C dx^A \wedge dx^B.
\label{CovmediumMeq}
\end{eqnarray}
In this form, (\ref{mediumMeq}) and (\ref{CovmediumMeq})
make sense on any 4-manifold: no further structure is required
\cite{Kottler,Cartan,Schroedinger,Dantzig,Post}.
One simply has two closed 2-forms $F$ and $H$.

\goodbreak

To proceed one needs to relate them by a constitutive relation.
In the linear case this is taken to be of the form
\begin{equation}
F_{ab} = \half \kappa _{ab}\,^{cd} H_{cd}
\end{equation}
where $ \kappa_{ab}\,^{cd} =-  \kappa_{ba}\,^{cd} = - \kappa _{ab}\,^{dc}$ and
where $\kappa _{ab}\,^{cd}$ does not depend on $F$ or $H$.
In  standard general relativity with its given  Lorentzian metric $\rg$
one takes
\begin{equation}
F= \star_\rg H
\end{equation}
where $\star_\rg$ denotes Hodge dual. Given a stationary Lorentzian metric $\rg$,
possibly flat but in non-Cartesian coordinates,
one may deduce the constitutive relation \cite{Tamm,Skrotskii,Plebanski}.
This idea is  at the core of the \emph{transformation optics}
approach to designing \emph{cloaking devices}. One picks
the Lorentzian metric whose null geodesics  one wishes light rays to
follow and reads off the  properties of the \emph{metamaterials}
that are required \cite{Leonhardt}.

However it is possible to reverse this logic and ask
what further properties are required of  $ \kappa_{ab}\,^{cd} $ so that
one may  determine from it  a  conformal equivalence class
of Lorentzian metrics $\rg$~? From this point of view one regards
light as fundamental and the spacetime metric $g$  as a derived concept
defined by the constitutive relation.
This program was initiated by Peres \cite{Peres} and
has been actively pursued by \cite{Hehl1,Hehl2}. In particular one may investigate the characteristic
\emph{wave surfaces} or their Legendre dual \emph{ray surfaces}
of the resulting  equations. In general these are given by a  quartic cone
and so do not define a Lorentzian structure for spacetime.
In the previous section we have seen how, in
what from this perspective are rather degenerate  cases,  Galilei and
and Carroll structures  can emerge for special choices of
the constitutive relation. An interesting question is whether
there exist metamaterials with these constitutive relations.

\subsection{Electric-Magnetic Duality}

The sourceless Maxwell equations  (\ref{mediumMeq}) can also be rewritten as
\begin{equation}
\vnabla \times {\vM} =- i \frac{\partial{\vN}}{\partial t} \,,\qquad
\vnabla \cdot \vN =0 \,,
\end{equation}
where
\begin{equation}
\vN= \vB+i\vD \,, \qquad \vM = \vH + i \vE \,.
\end{equation}
The \emph{constitutive relation}  may be expressed as
\begin{equation}
\vM=\vM(\vN) \,. \label{constitutive}
\end{equation}

An \emph{$SO(2)$ electric-magnetic duality rotation} is
the $SO(2)$ action :
\begin{equation}
\vN \rightarrow  e^{i\theta} \vN \,, \qquad  \vM \rightarrow  e^{i\theta} \vM
\label{rotation}
\end{equation}
and the question is whether the constitutive relation (\ref{constitutive})
is invariant under the $SO(2)$ electric-magnetic duality rotation
(\ref{rotation}).

A less restrictive demand is that the constitutive relation
 is invariant under the discrete involution
corresponding to $\theta= \frac{\pi}{2}$, i.e. under
\begin{equation}
\vB \rightarrow -\vD \,, \qquad \vD \rightarrow \vB \,,\qquad
\vH \rightarrow -\vE \,, \qquad \vE \rightarrow \vH \,. \label{discrete}
\end{equation}

One may check that the constitutive  relations
for  Carollian electrodynamics are not invariant under  (\ref{discrete}).
This is consistent with the results of \cite{BunsterHe}.

\section{Chaplygin gas}\label{Chaply}

Bazeia and Jackiw \cite{BaJa} pointed out that the \emph{non-relativitic system in $d$ space dimensions} called \emph{Chaplygin gas} carries a strange, \emph{field-dependent $(d,1)$-dimensional ``dynamical" Poincar\'e symmetry}.

Let us briefly outline how this comes about.
A rotation-free compressible fluid with density  $\rho$ and velocity $\bv=\bnabla\theta$ is described by the Euler equations,
\beqa
\p_t\rho+\vnabla\cdot(\rho\vnabla)=0,
\qquad
\p_t\theta+\half(\vnabla\theta)^2=-\frac{d V}{d\rho},
\label{EulerBernoulli}
\eeqa
where $V=V(\rho)$ is some potential. It is then straightforward to show that a Galilean boost in $1+1$ Galilei space-time,
$
x'=x+\beta{}t,\; t'=t,
$
implemented on the fields as
\beq
\rho'(x,t)=\rho(x',t'),
\qquad
\theta'(x,t)=\theta(x',t')-\beta{}x-\half\beta^2t,
\label{ChapImpl}
\eeq
leaves the equations of motion (\ref{EulerBernoulli}) invariant \footnote{Note that  (\ref{ChapImpl}) is precisely the way a boost acts on the phase of a wave function, $\psi=\rho\,e^{i\theta/\hbar}$, for a unit mass.
For simplicity, we only study the case $d=1$.}: the theory is Galilei-invariant, as expected.

\goodbreak

In the particular case when  the \emph{``Chaplygin'' potential $V\propto 1/\rho$ is chosen}, the system has more symmetries, though, namely
\begin{equation}
\begin{array}{ll}
\left\{\begin{array}{c}
{x}'=
x+\alpha\theta({x}',{t}')
\hfill
\\
{t}'=t+\frac{1}{2}\alpha
\big(x+{x}')
\\
\end{array}\right.\qquad
\hfill
&\hbox{``antiboost''}
\\[22pt]
\left\{\begin{array}{c}
{x}' =x
\hfill
\\
t'=e^{\delta}t\hfill
\\
\end{array}\right.
\hfill
&\hbox{time dilation}
\\
\end{array}
\label{BJtransf}
\end{equation}
with $\alpha,\delta\in\IR$. ``Antiboosts'' are particularly interesting:
 $x' $ and
${t}' $ are only defined implicitly,
and the action is ``field--\-depen\-dent''
in that, in addition to coordinates, its very definition  involves the field $\theta$.
Implementing them non--conventionally,
\begin{equation}
\begin{array}{ll}
\left\{\begin{array}{cll}
\rho'(x,t)&=&
\displaystyle{\frac{\rho({x}',{t}')}{J'}}\hfill
\\
\theta'&=&\theta({x}',{t}')
\qquad\hfill
\\
\end{array}\right.\hfill
&\hbox{``antiboost''}\hfill
\\[22pt]
\left\{\begin{array}{cll}
\rho'(x,t)&=&
e^{-\delta} \rho({x}',{t}')
\qquad\hfill
\\
\theta'(x,t)&=&
e^{\delta}\theta({x}',{t}')\hfill
\\
\end{array}\right.\hfill
&\hbox{time dilation}\hfill
\label{BJimp}
\end{array}
\end{equation}
where
$J'=
\Big[1-\alpha\partial_{{x}'}\theta({x}',{t}')
-\frac{1}{2}
\alpha^2\partial_{{t}'}\theta({x}',
{t}')\Big]^{-1}$
is the Jacobian of the space-time transformation.
Equations (\ref{BJtransf}) provide us with further symmetries. Even more intriguingly, combining the ``antiboosts'' and time dilations with those standard ones of centrally extended  Galilei yields a  Poincar\'e symmetry in $(2,1)$ dimensions.

Here we observe that, since the Carroll group is a subgroup of  Poincar\'e in one higher dimension, \emph{the Chaplygin gas  carries a Carroll symmetry} (but realized in a non-conventional way).

The mystery has been explained in Ref. \cite{HHAP}
 by using the Bargmann framework.
Let us first observe that,
for $t=0$, the  implementation (\ref{ChapImpl}) on the velocity potential field is that of a Carrollian boost, (\ref{carrboost}), when $\theta$ is traded for Carrollian time, $s$.
Then the idea is that the field $-\theta'$ should be promoted to become the ``vertical'' coordinate $s$.
Then the funny-looking actions (\ref{BJtransf}) lift to Bargmann space as
\begin{equation}
\begin{array}{cc}
\hbox{antiboost}:\qquad\hfill
&\left\{\begin{array}{lll}
{x}&=&
x-\alpha s,
\\
{t}&=&t+\alpha x-\frac{1}{2}\alpha^2s,
\hfill
\\
{s}&=&s,\hfill
\\
\end{array}\right.
\end{array}
\label{BJBtransf}
\end{equation}
which is precisely a \emph{Carroll boost} lifted to Bargman space -- which is now $(2,1)$-dimensional Minkowski space with light-cone coordinates $t$ and $s$. This should be compared with that of a lifted \emph{Galilean boosts}\begin{equation}
\begin{array}{cc}
\hbox{Galilei boost}:\qquad
\left\{\begin{array}{lll}
{x}&=&x+\beta t,
\\
{t}&=&t,
\\
{s}&=&s-\beta x-\frac{1}{2}\beta^2t.
\\
\end{array}\right.
\end{array}
\label{GalBtr}
\end{equation}
The two boost actions are obtained from each other by the ``duality'' interchange,
\begin{equation}
t\longleftrightarrow s
\label{tsinterchange}
\end{equation}
and $\beta\leftrightarrow-\alpha$.
Dilations of time alone in Eq. (\ref{BJtransf}),
lifted to Bargmann space become isometries there which in fact remain
dilations of time alone under $ t \leftrightarrow s $ interchange, but with the inverse parameter,
$\delta\to-\delta$, namely
\begin{equation}
\begin{array}{cccc}
\hbox{time dilations}:\;
&\left\{\begin{array}{lll}
{x}'&=&x,
\hfill
\\
{t}'&=&e^{\delta}t,
\\
{s}'&=&e^{-\delta}s,
\end{array}\right.
&\Longleftrightarrow\quad\quad
&\left\{\begin{array}{lll}
{x}&=&x,
\\
{t}&=&e^{-\delta}t,
\\
{s}&=&e^{\delta}s.
\end{array}\right.
\end{array}
\label{BJBtransf2}
\end{equation}
The same rule changes a time translation into a ``phase translation'',
\beq
\left\{\begin{array}{cll}
{x}'&=&x,
\\
{t}'&=&t+\epsilon,
\\
{s}&=&s,
\end{array}\right.
\qquad\Longleftrightarrow\qquad
\left\{\begin{array}{cll}
{x}'&=&x,
\\
{t}'&=&t,
\\
{s}'&=&s-\eta.
\end{array}\right.
\label{tvtr}
\end{equation}
Augmented with ordinary space translations, our transformations span the the isometries of
$(2,1)$ dimensional Minkowski space, --
Poincar\'e group in $(2,1)$ dimensions, with the
Galilei and Carroll subgroups, interchanged by ``duality" (\ref{tsinterchange}).

Lift, at last, the fields to Bargmann space according to
\beq
\widehat{\rho}(x,t,s)=\rho(x,t),
\qquad
\widehat{\theta}(x,t,s)=\theta(x,t)+s.
\eeq
Then the \emph{natural geometric action of the Poincar\'e group} turns out to be a symmetry for the lifted system.
Moreover, the action ``upstairs'' of the non-Galilei generators reduces to the ``funny ones'' downstairs \cite{HHAP}.

 \section{Conclusion}

The aim of this paper has been to point out the
fascinating duality between the usual  Galilean and L\'evy-Leblond's  more subtle ``Carrollian'' limits. Both limits are obtained by  Wigner-In\"on\"u  \cite{Inonu} contraction of the Poincar\'e group when a suitable parameter, $c$ and $C$, respectively, goes to infinity.

Both limits can be considered as applied to electromagnetism:
the first yields the two kinds of Galilean electromagnetism as put forward by Le Bellac and L\'evy-Leblond \cite{LBLL},
while the other one yields two kinds of Carroll-invariant ``electromagnetisms'' \footnote{The full Maxwell equations are invariant under the relativistic conformal group $\Ort(4,2)$ and their ``magnetic-type'' NR limit has been shown to carry a Conformal Galilei symmetry \cite{NOR,DHGalConf}.}.

In vacuum, suitably redefined (Maxwellian) electromagnetic fields satisfy a
wave equation with propagation speed $c$ and $C^{-1}$, respectively. The Galilean limit arises hence when \emph{the velocity of light, $c$, --- measured in Newton's time, $t$}, --- goes to \emph{infinity}, whereas the ``Carrollian limit'' is
one when \emph{the velocity of light, $C^{-1}$,  --- but one measured in ``Carrollian time'', $s$}, goes to \emph{zero}.
Their intuitive meaning is that, in the Galilean case, the light-cone ``umbrella'' opens up to become a spacelike slice $t=\const$, while in the Carrollian limit, it collapses to a timelike-axis parametrized by $s$ \cite{Leblond}.

\goodbreak

The two limits can be unified by lifting it to relativistic ``Bargmann'' space, which also  unifies the Galilei and Carroll groups.

We just mention that the above-mentioned duality hints at  various \emph{conformal extensions of the Carroll group}, analogous to \emph{Conformal Galilei groups}.
Recent work \cite{Bag2} hints, for example, at an intriguing
relation between the Bondi-Metzner-Sachs (BMS)
\cite{BMS} and the Conformal Galilei (CG) groups.
The BMS group is,
in fact, the \emph{conformal extension of the Carroll Group}.  Details will be published separately \cite{DGHunpub}.

Can \emph{both} Galilean and Carrollian symmetry coexist for the same physical system~? Such an example is
 provided by the \emph{Chaplygin gas}, whose
 Poincar\'e symmetry,  is indeed the isometry of the unifying Bargmann space, discussed in Section \ref{Unifchap}.

One can also wonder whether \textit{bona fide} particles with Carroll symmetry do exist. Then answer is \emph{yes} --- but
they have a rather limited interest: they cannot move!
The proof is outlined in the Appendix; see also Ref. \cite{Gomis,Ancille}.


\begin{acknowledgments}
G.W.G would like to thank the {\it Galileo Galilei Institute}  in
Florence for its hospitality during the 2007 \textit{Workhop on
String and M-Theory Approaches to Particle Physics and Cosmology},
and for many illuminating  conversations with Quim Gomis on the
Carroll group at that time, and  {\it KITP} in Santa Barbara for its
hospitality during its 2012 \textit{Bits and Branes workshop}
discussions at which stimulated the present investigation. P.A.H is
indebted to the \emph{Institute of Modern Physics} of the Lanzhou
branch of the Chinese Academy of Sciences for hospitality. This work
was partially supported by the National Natural Science Foundation
of China (Grant No. 11035006 and 11175215) and by the Chinese
Academy of Sciences Visiting Professorship for Senior International
Scientists (Grant No. 2010T1J06).
\end{acknowledgments}



\newpage

\appendix{\bf Appendix: Carrollian particles}

\renewcommand{\theequation}{A\thesection.\arabic{equation}}
\renewcommand
\appendix{\appendix}{\setcounter{equation}{0}}

``Elementary'' particles associated with a given space-time symetry group $G$ can conveniently be constructed by Souriau's method \cite{Sou}.
The general construction starts with the Lie group $G$ with Lie algebra $\fg$ and a point $\mu_0\in\fg^*$ that serves as the origin of the coadjoint orbit $\cO_{\mu_0}=\Coad(G)\mu_0$. The exterior derivative of the real $1$-form
$
\varpi=\mu_0\cdot\Theta
$
 where $\Theta=``g^{-1}dg''$ is the (left-invariant) Maurer-Cartan $1$-form of $G$, $\sigma=d\varpi$, descends as the canonical KKS symplectic $2$-form, $\omega$ of $\cO_{\mu_0}$, namely $\sigma=(G\to\cO_{\mu_0})^*\omega$.

In flat space, the Carroll group has no symplectic cohomology \cite{LLcoho} if $d\geq3$.
(For the planar case, see \cite{Ancille}). The ``space of motions'' \cite{Sou} of a \emph{Carrollian elementary particle}, constructed as homogeneous symplectic spaces, is therefore simply a coadjoint orbit of $\Carr(d+1)$ itself.


The group $\Carr(d+1)$ is
represented by the matrices $a$ in (\ref{Carr}), and its Lie algebra $\carr(d+1)$ given  in (\ref{carralg}) or in (\ref{Z}).
Then an element of the dual of the Lie algebra retains the form
$
\mu=(\bsell,\bg,\bp,m)\in\carr(d+1)^*
$,
where the pairing between the Lie algebra and its dual is defined by
\begin{equation}
\mu\cdot{}Z=\half\Tr(\bsell\bomega)-\bg\cdot\bbeta-\bp\cdot\bgamma
+m\varphi,
\label{muZ}
\end{equation}
for all $Z\in \carr(d+1)$.
The coadjoint action, deduced from the adjoint one, is given by $\Coad(a)\mu\equiv\mu\circ\Ad(a^{-1})$, reads  $\Coad(a)(\bsell,\bg,\bp,m)=(\bsell',\bg',\bp',m')$, where
\begin{eqnarray}
\label{ell}
\bsell'
&=&
R\bsell{}R^{-1}+(R\bg\,\bb^T-\bb\,(R\bg)^T)+(R\bp\,\bc^T-\bc\,(R\bp)^T)+m(\bc\,\bb^T-\bb\,\bc^T)\\[4pt]
\label{g}
\bg'
&=&
R\bg+m\bc\\[4pt]
\label{p}
\bp'
&=&
R\bp-m\bb\\[4pt]
\label{m}
m'
&=&
m
\label{Coad}
\end{eqnarray}
showing that $m$ is a Casimir invariant, readily interpreted as the \emph{mass}.

\begin{itemize}
\item
If $m\neq0$, define
\beq
\bs=\bsell+\frac1m(\bg\,\bp^T-\bp\,\bg^T)
\eeq
 so that $\bs'=R\bs{}R^{-1}$, which yields another Casimir invariant, viz., the (scalar) \textit{spin} $\mathsf{s}$,
\begin{equation}
\mathsf{s}^2=-\half\Tr(\bs^2)
\label{s2}
\end{equation}

\item
If $m=0$, we find three extra invariants, namely
\begin{equation}
p=\Vert\bp\Vert
\qquad
\&
\qquad
g=\Vert\bg\Vert
\qquad
\&
\qquad
w=\bg\cdot\bp
\label{p2g2pg}
\end{equation}
The invariant $p$ is indeed reminiscent of the Euclidean coadjoint invariant Souriau calls ``color''~\cite{Sou}, with the same physical dimension as the Minkowskian Pauli-Lubanski vector. Also, $w$ is in turn analogous to the ``helicity''.  These Carroll invariants have physical dimension $[p]=AL^{-1}$, $[g]=ML$, and $[w]=MA$ where $[A]=[\hbar]$.
\end{itemize}


Let us deal with, e.g., \emph{spinless massive} free Carrollian particles by choosing
$
\mu_0=(0,0,0,m)
$
with $m>0$. Then the associated $1$-form reads
\begin{equation}
\varpi=m\,\delta_{AB}\,v^A dx^B + m\,ds,
\label{varpiBis}
\end{equation}
whose exterior derivative clearly descends to the ``evolution'' space $V=T\bbR^d\times\bbR\ni(\bx,\bv,s)$,  endowed with the presymplectic 2-form,
\begin{equation}
\sigma=m\,\delta_{AB}\,dv^A\wedge{}dx^B.
\label{sigma}
\end{equation}
The ``equations of motion'' are given by the characteristic foliation $\ker\sigma$,
whose integration yields a \emph{desperately poor ``dynamics''} for free massive Carrollian particles, viz.,
\begin{equation}
\bx(s)=\bx(0),
\qquad
\bv(s)=\bv(0),
\label{freemotions}
\end{equation}
for all $s\in\IR$. The associated space of ``motions'' is therefore $(T^*\bbR^d,\omega)$ with $\omega=dp_A\wedge{}dq^A$ where $\bp=m\vv$ and $\bq=\bx$.

 From the Bargmannian point of view, these curves are the restrictions to the Carroll manifold~$C$  of the null geodesics of the Bargmann manifold  whose velocity is orthogonal to~the null vector field~$\xi$ --- and  hence
 ``vertical'', i.e., parallel to $\xi$.

In conclusion, the \textit{Red Queen} was right: \textit{even running very fast, one does not advance in the Carroll World!}


\begin{thebibliography}{99}

\bibitem{Leblond} J. M. L\'evy-Leblond,
``Une nouvelle limite
non-relativiste du group de Poincar\'e,''
Ann. Inst. H. Poincar\'e {\bf 3} (1965) 1. 

\bibitem{Inonu}
  E.~In\"on\"u and E.~P.~Wigner,
``On the contraction of groups and their representations,''
  Proc.\ Nat.\ Acad.\ Sci.\  {\bf 39} (1953) 510.

\bibitem{SenGupta}
 V. D. Sen Gupta,
``On an Analogue of the Galileo Group,''
Il Nuovo Cimento {\bf 44} (1966) 512. 

\bibitem{Alice}
Lewis Carroll, \textit{Through the Looking Glass and what Alice Found There}.
 London: MacMillan  (1871).


\bibitem{Newton}
I.~S.  Newton, {\it Philosophia
Naturalis Pricipia Mathematica} London: Royal Society of London
(1686), translated by A.~ Motte
as  {\it Sir
Isaac Newton's Mathematical Principles of Natural Philosphy
and his System of the World} (1729). Translation revised
by F.~Cajori, Berkeley: University of Caliornia Press (1946).
\bibitem{DBKP}
  C.~Duval, G.~Burdet, H.~P.~K\"unzle and M.~Perrin,
  ``Bargmann Structures and Newton-Cartan Theory,''
  Phys.\ Rev.\ D {\bf 31} (1985) 1841.

\bibitem{DGH}
C. Duval, G.W. Gibbons, P. Horvathy,
``Celestial mechanics, conformal structures and gravitational waves,''
 Phys. Rev. {\bf D43} (1991) 3907 
[hep-th/0512188].


\bibitem{HHAP}
M. Hassa\"\i ne and P. A. Horv\'athy,
``Field--dependent symmetries of a non-relativistic fluid model,''
Ann. Phys. (N. Y.)  {\bf 282}, 218 (2000)
 [math-ph/9904022].

\bibitem{DLazzari}
  C.~Duval and S.~Lazzarini,
``Schr\"odinger Manifolds,''
  J.\ Phys.\ A {\bf 45} (2012) 395203
  [arXiv:1201.0683 [math-ph]].

\bibitem{Bag2}
  A.~Bagchi,
  ``Correspondence between Asymptotically Flat Spacetimes and Nonrelativistic Conformal Field Theories,''
  Phys.\ Rev.\ Lett.\  {\bf 105} (2010) 171601.

\bibitem{Barnich1}
  G.~Barnich and C.~Troessaert,
 ``Aspects of the BMS/CFT correspondence,''
  JHEP {\bf 1005} (2010) 062
  [arXiv:1001.1541 [hep-th]];
 ``BMS charge algebra,''
  JHEP {\bf 1112} (2011) 105.

\bibitem{Arcioni}
  G.~Arcioni and C.~Dappiaggi,
``Holography in asymptotically flat space-times and the BMS group,''
  Class.\ Quant.\ Grav.\  {\bf 21} (2004) 5655
  [hep-th/0312186].

\bibitem{Schroer}
  B.~Schroer,
``Bondi-Metzner-Sachs symmetry, holography on null-surfaces and
area proportionality of `light-slice' entropy,''
  Found.\ Phys.\  {\bf 41} (2011) 204
  [arXiv:0905.4435 [hep-th]].

\bibitem{Henneaux1}
M.~Henneaux,
 ``Geometry Of Zero Signature Space-times,''
  Bull.\ Soc.\ Math.\ Belg.\  {\bf 31} (1979) 47;
M. Henneaux,
Acad. Roy. Belg., Bull. Sci. (5)
 {\bf 68} (1982) 940. 

\bibitem{Melas}
  E.~Melas,
``Open problems and results in the group theoretic approach to quantum gravity via the BMS group and its generalizations,''
  J.\ Phys.\ Conf.\ Ser.\  {\bf 283} (2011) 012023.

\bibitem{thoughts}
  G.~W.~Gibbons,
 ``Thoughts on tachyon cosmology,''
  Class.\ Quant.\ Grav.\  {\bf 20} (2003) S321
  [hep-th/0301117].

 \bibitem{GiHaYi}
 G. W. Gibbons, K. Hashimoto and Piljin Yi,
``Tachyon condensates, Carrollian contractions of the Lorentz group and fundamental strings,''
JHEP (2002) 0209: 061 [hep-th/0209034].

 \bibitem{Max} G. W. Gibbons,
``The Maximum Tension Principle in General
Relativity,''
Foundations of Physics   [hep-th/021009].

 \bibitem{Dautcourt}
  G.~Dautcourt,
 ``On the ultrarelativistic limit of general relativity,''
  Acta Phys.\ Polon.\ B {\bf 29} (1998) 1047
  [gr-qc/9801093].

  \bibitem{Nicolai} T. Damour, M. Henneaux and H. Nicolai,
``Cosmological Billiards,''
   [hep-th/0212256].

\bibitem{Wall1}
  A.~C.~Wall,
``A Proof of the generalized second law for rapidly-evolving Rindler horizons,''
  Phys.\ Rev.\ D {\bf 82} (2010) 124019
  [arXiv:1007.1493 [gr-qc]].

\bibitem{Wall2}
  A.~C.~Wall,
``A proof of the generalized second law for rapidly changing fields and arbitrary horizon slices,''
  Phys.\ Rev.\ D {\bf 85} (2012) 104049
  [arXiv:1105.3445 [gr-qc]].

\bibitem{BunsterHe}
  C.~Bunster and M.~Henneaux,
 ``Duality invariance implies Poincare invariance,''
  Phys.\ Rev.\ Lett.\  {\bf 110} (2013) 011603
  [arXiv:1208.6302 [hep-th]].

\bibitem{GalIso}
C.~Duval, ``Galilean isometries,''
Class. Quantum Grav. {\bf 10} (1993), 2217
[arXiv:0903.1641].

\bibitem{DHGalConf}
C.~Duval, and P.~A.~Horvathy
``Non-relativistic conformal symmetries and Newton-Cartan structures,''
J. Phys. A {\bf 42}  (2009) 465206;
[arXiv:0904.0531];
  ``Conformal Galilei groups, Veronese curves, and Newton-Hooke spacetimes,''
 J. Phys. {\bf A 44} (2011) 335203
 [arXiv:1104.1502].

\bibitem{LBLL}
 M. Le Bellac and J.-M. L\'evy-Leblond,
``Galilean Electromagnetism,''
 Nuovo Cimento {\bf 14B} (1973), 217

\bibitem{Kun2}
H.\ P.\ K\"unzle,
``Covariant Newtonian Limit of Lorentz Space-Times,''
Gen.\ Relativ.\ Gravit.\ {\bf 7}, 445 (1976).

\bibitem{SourElec}
J.-M. Souriau,
``Galilean Electrodynamics'' (In French)
CPT-85/PE-1831 (1985).

\bibitem{GoSh}
G. A. Goldin and V. M. Shtelen,
``On Galilean invariance and nonlinearity in electrodynamics and quantum mechanics,''
Phys. Lett. {\bf A279} (2001) 321.

\bibitem{Rousseaux}
  J.~M.~Houlrik and G.~Rousseaux,
 ``'Nonrelativistic' kinematics: Particles or waves?''
  arXiv:1005.1762 [physics.gen-ph].

\bibitem{BaJa}
D.~Bazeia and R.~Jackiw,
 ``Nonlinear realization of a dynamical Poincar\'e symmetry by a field-dependent diffeomorphism,''
  Annals Phys.\  {\bf 270} (1998) 246;
  R.~Jackiw,
 ``A Particle field theorist's lectures on supersymmetric, nonAbelian fluid mechanics and d-branes,''  physics/0010042.

\bibitem{Gomis}
J. Gomis and F. Passerini, ``Super Carroll space,
Carrollian super-particle and Carrollian super-string,'' unpublished notes (2005)

\bibitem{Ancille}
  A.~Ngendakumana, J.~Nzotungicimpaye and L.~Todjihounde,
  ``Group theoretical construction of planar Noncommutative Phase Spaces,''
  J. Math. Phys. {\bf 55}, 013508 (2014). %
  [arXiv:1308.3065 [math-ph]]. See also
A.~Ngendakumana, ``Group theoretical construction of planar noncommutative systems,'' Doctoral Thesis at the \textit{Institut de Math\'ematiques et de Sciences Physiques de l'Universit\'e d'Abomey-Calavi, Porto Novo (Benin)}, (2013). [arXiv:1401.5213 [math-ph]]

\bibitem{Huang}
  H.~-Y.~Guo, C.~-G.~Huang, H.~-T.~Wu and B.~Zhou,
 ``The Principle of Relativity, Kinematics and Algebraic Relations,''
  Sci.\ China G {\bf 53} (2010) 591
  [arXiv:0812.0871 [hep-th]];
  C.~-G.~Huang, Y.~Tian, X.~-N.~Wu, Z.~Xu and B.~Zhou,
 ``Geometries for Possible Kinematics,''
  Sci.\ China G {\bf 55} (2012) 1978
  [arXiv:1007.3618 [math-ph]].

\bibitem{DK}
C.~Duval and H.P.~K\"unzle,
``Sur les connexions newtoniennes et l'extension
non triviale du groupe de Galil\'ee,''
C.R. Acad. Sci. Paris
{\bf 285 A} (1977), 813. 

\bibitem{Kunzle}
H. P. K\"unzle, ``Galilei and Lorentz structures on space-time:
  Comparison of the corresponding geometry and physics'',
Ann. Inst. H. Poincar{\'e}. Phys. Th{\'e}or. {\bf 17} (1972) 337--362.

\bibitem{LLcoho}
J.-M.~L\'evy-Leblond,
in {\it Group Theory and Applications}, Loebl Ed.,
{\bf II}, Acad. Press, New York, p.~222 (1972).

\bibitem{Sou}
J.-M.~Souriau,
\textsl{Structure des syst\`emes dynamiques}, Dunod (1970, \copyright 1969);
\textsl{Structure of Dynamical Systems. A Symplectic View of Physics},
Birkh\"auser (1997).

\bibitem{Trautman}
A.~Trautman,
``Sur la th\'eorie newtonienne de la gravitation,''
C.R. Acad. Sci. Paris {\bf 257}
(1963),
617--620;
``Comparison of Newtonian and relativistic theories of
space time,'' pp.~413--425 in \textit{Perspectives in Geometry and
Relativity},   (B.~Hoffmann, ed.),
Indiana University Press, Bloomington, 1964.

\bibitem{DGHunpub}
C. Duval, G. Gibbons and P. A. Horvathy,
``Conformal Carroll groups and BMS symmetry'',
[arXiv:1402.5894 [gr-qc]].

\bibitem{Nature}
 L.V. Hau, S.E. Harris, Z. Dutton,
 Z. H. Behroozi,
 Nature {\bf 397}, 594 (1999);
  J. Marangos,
 ``Slow light in cool atom,''
Nature, {\bf 397} 559   (1999)
[18 FEBRUARY 1999, www.nature.com];
``Faster than a speeding photon'',
Nature, {\bf 406},  243  (2000)
[20 JULY 2000, www.nature.com].

\bibitem{NOR} J. Negro, M.~A. del Olmo, An A. Rodr\'{\i}guez-Marco,
``Nonrelativistic conformal groups,''
J. Math. Phys. {\bf 38} (1997), 3786, and
``Nonrelativistic conformal groups. II. Further developments and physical applications,''
\textit{ibid.} 3810.

\bibitem{BMS}
H. Bondi, M. G. van der Burg, and A. W. Metzner,
``Gravitational waves in general relativity. 7.
Waves from axisymmetric isolated systems,''
Proc. Roy. Soc. Lond. A {\bf 269} (1962) 21;
  R.~Sachs,
  ``Asymptotic symmetries in gravitational theory,''
  Phys.\ Rev.\  {\bf 128} (1962) 2851.

\bibitem{Kottler} F. Kottler, Sitzungsber. Akad. Wien IIa 131, 119 (1922).

\bibitem{Cartan}  E. Cartan, ``On manifolds with an affine connection and the theory of general relativity'' (Bibliopolis, Napoli, 1986), english
translation of the french original from 1923/24.

\bibitem{Dantzig}
 D.~V.~ Dantzig, Proc. Cambridge Philos. Soc. 30, 421 (1934).

\bibitem{Schroedinger}  E. Schroedinger,
``Space-time structure'' (Cambridge University Press, Cambridge, 1950).

\bibitem{Post} E.~ J.~Post,
``Formal structure of electromagnetics,'' (North-Holland, Amsterdam, 1962).

\bibitem{Tamm}
I.~ E.~ Tamm, Zh. Rus. Fiz.-Khim. Obshchestva, Otd. Fiz.
{\bf 56}, 248 (1924).

\bibitem{Skrotskii}
 G.~V.~ Skrotskii, Dokl. Akad. Nauk SSSR {\bf 114}, 73
 (1957) [Soviet Physics Doklady 2, 226 (1957)]

\bibitem{Plebanski}
J.~Plebanski,
``Electromagnetic Waves in Gravitational Fields,''
  Phys.\ Rev.\  {\bf 118} (1959) 1396.

\bibitem{Leonhardt} U. Leonhardt and T. G. Philbin,
``Transformation Optics and the Geometry of Light,''
 {\it Prog. Opt.} {\bf 53}(2009) 69-152
[arXiv:0805.4778].

\bibitem{Peres}
A. Peres,
``Electromagnetism, geometry, and the equivalence principle,'' {\it Ann. Phys.}
{\bf  19} (1962) 279-286.

\bibitem{Hehl1}F. W. Hehl and Y. Obukhov, ``Foundations of
 classical electrodynamics,''
 (Birkhauser, Basel, 2003).

\bibitem{Hehl2}
  F.~W.~Hehl and Y.~N.~Obukhov,
``To consider the electromagnetic field as fundamental, and the metric only as a subsidiary field,''
Found.\ Phys.\  {\bf 35} (2005) 2007
  [physics/0404101].

\end{thebibliography}
\end{document}